\documentstyle[eqsecnum,floats,prd,aps,epsf]{revtex}

\def\bm{\boldmath}

\def\T{{\cal T}}
\def\R{{\cal R}}
\def\gr{{\rm field}}
\def\ma{{\rm particle}}
\def\imax{{i_{\rm max}}}

\def\la{\lambda}

\def\p{\partial}

\def\vt{\vartheta}
\def\vp{\varphi}

\begin{document}
\twocolumn[\hsize\textwidth\columnwidth\hsize\csname@twocolumnfalse\endcsname

\title{Gravitational Wavetrains in the
Quasi-Equilibrium Approximation:\\ 
A Model Problem in Scalar Gravitation}

\author{Hwei-Jang Yo$^{1,2}$, Thomas W. Baumgarte$^1$, and
	Stuart L. Shapiro$^{1,3}$} 

\address{$^1$Department of Physics,
University of Illinois at Urbana-Champaign, Urbana, Il 61801\\ 
$^2$Institute of Astronomy and Astrophysics, Academia Sinica, 
Taipei 115, Taiwan, ROC\\ 
$^3$Department of Astronomy \& NCSA, University of Illinois at
Urbana-Champaign,Urbana, Il 61801} 

\maketitle

\begin{abstract}
A quasi-equilibrium (QE) computational scheme was recently developed
in general relativity to calculate the complete gravitational
wavetrain emitted during the inspiral phase of compact
binaries.  The QE method exploits the fact that the the
gravitational radiation inspiral timescale is much longer than the
orbital period everywhere outside the ISCO. Here we demonstrate the
validity and advantages of the QE scheme by solving a model problem in
relativistic scalar gravitation theory.  By adopting scalar
gravitation, we are able to numerically track without approximation
the damping of a simple, quasi-periodic radiating system (an
oscillating spherical matter shell) to final equilibrium, and then use
the exact numerical results to calibrate the QE approximation
method. In particular, we calculate the emitted gravitational
wavetrain three different ways: by integrating the exact coupled
dynamical field and matter equations, by using the scalar-wave
monopole approximation formula (corresponding to the quadrupole
formula in general relativity), and by adopting the QE scheme.  We
find that the monopole formula works well for weak field cases, but
fails when the fields become even moderately strong. By contrast, the
QE scheme remains quite reliable for moderately strong fields, and
begins to breakdown only for ultra-strong fields.  The QE scheme thus
provides a promising technique to construct the complete wavetrain
from binary inspiral outside the ISCO, where the gravitational fields
are strong, but where the computational resources required to follow
the system for more than a few orbits by direct numerical integration
of the exact equations are prohibitive.
\end{abstract}
\vskip2pc]

\section{Introduction}
The construction of several new gravitational wave detectors, including
the Laser Interferometer Gravitational Wave Observatory (LIGO), TAMA,
VIRGO and GEO, may soon make gravitational wave astronomy a reality.
The inspiral and coalescence of compact binaries, consisting of
neutron stars or black holes, are among the most promising sources for
detection by these observatories.  It is expected that neutron
star/neutron star binaries will spend approximately 16,000 cycles in
the LIGO/VIRGO frequency band, neutron star/black hole binaries about
3,500, and black hole/black hole binaries about 600~\cite{kip}.  To
increase the likelihood of detection, and to extract information from
the signal, the binary inspiral has to be modeled theoretically and
waveform templates have to be constructed.

The evolution of compact binaries proceeds in different stages.  By
far the longest is the initial quasi-equilibrium inspiral stage,
during which the compact objects move in nearly circular orbits, while
the separation between them slowly decreases as energy is carried away
by gravitational radiation.  The quasi-circular orbits become unstable
at the innermost stable circular orbit (ISCO), where the inspiral
enters a plunge and merger phase.  The merger and coalescence happens
on a dynamical timescale, and produces either a black hole or, for
binary neutron stars, possibly a larger neutron star, which may
collapse to a black hole at a later time.  The final stage of the
evolution is the ringdown phase, during which the merged object
settles down to equilibrium.

Two distinct approaches have commonly been employed to analyze the
adiabatic inspiral phase.  Much progress has been made in
post-Newtonian studies of compact binaries (see e.g.~\cite{djs00} and
references therein).  Most of these approaches, however, approximate
the compact objects as point sources, which neglects important
finite-size effects which may be particularly important
for neutron star binaries.  Also,
PN expansions may not converge sufficiently rapidly in the
strong-field region near the ISCO.  Alternatively, compact binaries
can be modeled numerically.  Computational constraints currently limit
dynamical evolution calculations to at most a few orbits, so that
there is no hope to simulate a complete inspiral.  It is possible,
however, to numerically model binaries in the adiabatic inspiral
phase in a quasi-equilibrium (QE) approximation.

The QE approximation is based on the assumption that the
gravitational radiation reaction time scale is much longer than the
orbital timescale, so that the binary can be approximated to be in
quasi-equilibrium (and quasi-circular orbit) on an orbital timescale.
A similar approximation is routinely being used in stellar evolution
calculations.  There, the evolutionary timescale is much longer than
the hydrodynamical timescale, so that the star can safely be assumed
to be in quasi-equilibrium on a dynamical timescale.
Quasi-equilibrium models of compact binaries have been constructed
both for neutron stars (see, e.g.,~\cite{bcsst} for corotating
and~\cite{irrotational} for irrotational binary neutron stars) and for
black holes~\cite{c94}.

Even though individual QE models only represent ``snap-shots'' of
binaries at a certain separation, it is possible to construct the
complete adiabatic inspiral together with the emitted gravitational
wave signal using the following scheme~\cite{stu}.  For each
separation, a QE model can be inserted as a matter source into
Einstein's equations (this is the ``hydro-without-hydro'' approach
demonstrated in~\cite{bhs}).  Numerically integrating Einstein's
equations for this given matter source will then yield the
gravitational wave signal and luminosity of the binary at that
separation.  Interpolation between a discrete sample of separations
then yields the gravitational wave luminosity as a continuous function
of separation.  Combining this with the binary's binding energy as a
function of separation, one can construct the inspiral rate at all
separations, and hence the separation as a function of time.  Given a
suitable parameterization, the entire continuous inspiral wavetrain can
then be constructed.  The viability of this approach has recently been
demonstrated in~\cite{dbs} where a prototype calculation was presented
for corotating binary neutron stars obeying a polytropic equation of
state.  Note that unlike post-Newtonian approaches, the QE scheme uses
the fully nonlinear Einstein equations.  The QE approach is also
computationally very efficient, since it is possible to produce a very
large number of cycles from a small number of QE configurations, each
of which needs to be followed for only a couple of orbital periods.

In this paper, we evaluate the QE approach for a model problem in
relativistic scalar gravitation, and show that it produces excellent
agreement with an exact numerical solution.  While scalar gravity
(see, e.g., problem 7.1 in Misner, Thorne and Wheeler~\cite{mtw}) is
not the correct theory of gravitation, it is conceptionally much
simpler than general relativity (GR), and shares many of its
characteristic features.  This makes scalar gravity a very attractive
framework for calibrating the QE approximation scheme, since, unlike
in GR, we can directly compare its results with the readily producable
exact solution.

In scalar gravity, the gravitational field is described by one scalar
function.  Also, scalar gravity generates gravitational radiation even
in spherical symmetry, so that the generation and emission of
gravitational waves can be studied with a spherically symmetric
numerical code, involving only one spatial coordinate.  This enables
us to track the effect of radiation reaction exactly over many
dynamical timescales.  Moreover, outgoing wave boundary conditions can
be imposed correctly at arbitrarily close separations from the sources
in spherical symmetry, which eliminates the need for large
computational grids.  The theory also admits a local law of energy
conservation, while GR only obeys global energy conservation.  In
numerical work, such a conservation law provides a strong check on the
accuracy of integration.  For all of these reasons, scalar gravitation
has been employed successfully to develop many tools of numerical
relativity (see, e.g., Shapiro and Teukolsky~\cite{slst}, hereafter
ST, and also~\cite{slst2}) and we extend that tradition here.

In this paper we study, in the framework of scalar gravitation, the
damped oscillations of a spherical matter shell, which we adopt as a
simple spherically symmetric analogue to binary inspiral in GR.  A
relativistic binary emits gravitational waves, which slowly extracts
energy from the binary orbit, so that the inspiral proceeds along a
sequence of nearly periodic circular orbits of decreasing separation.
Similarly, an oscillating matter shell emits gravitational waves,
which slowly extracts energy from the oscillation, so that the damping
proceeds along a sequence of nearly periodic oscillations of
decreasing amplitude.  Similar to relativistic binaries, where we
consider the quasi-circular orbits at a certain separation of the
adiabatic inspiral to be in QE, we may also consider the matter
shell's quasi-periodic oscillations of a certain amplitude to be in
QE.  One difference between the two processes is that the binary
inspiral continues until coalescence and merger, while the damping of
the matter shell's oscillations will continuously slow down until a
true equilibrium state has been reached.

We adopt three distinct approaches to computing the damped
oscillation of the matter shell.  We first compute the exact solution,
by numerically integrating the exact equations.  
In our second approach, we adopt
a QE approximation by neglecting gravitational waves, and computing QE
models of the periodic, undamped oscillations of matter shells for various
oscillation amplitudes.  These QE models are then inserted 
as matter sources into the dynamical equations for the gravitational
field.  Integrating the field equations for these QE matter sources
yields the gravitational wave form and luminosity for each oscillation
amplitude. Combining the wave luminosity with the QE oscillation energy as a
function of amplitude yields the amplitude decay rate together with the
continuous gravitational wavetrain.  
This is the analogue to the QE approach to
binary inspiral as outlined above.  In a third approach, we use a
monopole formula to compute the gravitational waveforms and 
wave luminosity for QE
models and hence to determine the inspiral rate.  This is the
equivalent to using the quadrupole formula in GR to compute the
inspiral rate and waveforms for binary models.

We compare these three different approaches for three different
initial configurations, representing weak, moderately strong and
ultra-strong field cases.  We find that for the weak field
configuration, all three approaches agree very well.  For the
moderately strong field configuration, the QE approach still agrees
very well with the exact solution, while the agreement with the
monopole result is much worse.  Only for ultra-strong fields, for
which the assumptions of QE break down, do we find a disagreement
between the QE approach and the exact solution.  However, even in this
case, the break-down is gradual and not abrupt.  This is a very
encouraging result, and suggests that the QE approach is a very
reliable and efficient framework for computing adiabatic binary
inspiral up to the ISCO.

The paper is organized as follows.  We summarize the basic equations
of scalar gravity in Sec.~II, and outline the QE scheme in the
Sec.~\ref{QES}.  In Sec.~\ref{numres} we present our numerical
results, and compare our three approaches for the three different
situations.  We summarize and discuss the implications of our findings
in Sec.~\ref{conc}.  We also include three Appendices.  Appendix A
contains a short proof that a minimum in the quasi-equilibrium energy
corresponds to a static equilibrium shell solution. Appendix B
presents some details of the QE scheme for constructing a continuous
wavetrain. Appendix C describes our numerical implementation of the
field equation.  Throughout the paper we adopt gravitational units
with $G=c=1$.


\section{The basic equations}\label{basic}


\subsection{Dynamical equations}

We follow exercise (7.1) in Misner, Thorne and Whee\-ler~\cite{mtw}
and study test particles in a relativistic scalar gravitational theory
(see also ST).  The field equation for the scalar gravitational field
$\Phi$ is
\begin{equation}\label{fe}
\Box\Phi=4\pi e^\Phi\rho,
\end{equation}
where the metric is the flat Minkowski metric
\begin{equation}
g_{\alpha\beta} = \eta_{\alpha\beta}^{\rm flat}.
\end{equation}
Note that the exponential term on the right hand side makes the field
equation nonlinear.  For a single particle of rest mass $m$, traveling
along its worldline $z^\mu(t)$, the 
comoving density $\rho$ can be written 
\begin{equation}
\rho=m{\delta^3[x^a-z^a(t)]\over\gamma\sqrt{-g}}={\rho_0\over\gamma},
\end{equation}
where $\gamma=\dot{z}^0$ is the Lorentz factor and $\rho_0$ is the
density in the stationary frame.  The particle follows a geodesic
\begin{equation}\label{pe}
{Du^\mu\over d\tau}+[g^{\mu\nu}+u^\mu u^\nu]\Phi_{,\nu}=0,
\end{equation}
where $D$ denotes covariant differentiation and $u^\mu=dz^\mu/d\tau$
is the 4-velocity.

Conservation of energy-momentum follows from
\begin{equation}\label{enmoconserv}
 \nabla\cdot\mbox{\boldmath$T$}=0,
\end{equation}
where the stress-energy tensor $T^{\mu\nu}$ 
can be decomposed into a gravitational field part and a matter part:
\begin{equation}
T_{\mu\nu}=T^{\gr}_{\mu\nu}+T^{\ma}_{\mu\nu},
\end{equation}
where
\begin{eqnarray}
&&T^\gr_{\mu\nu}={1\over4\pi}[\Phi_{,\mu}\Phi_{,\nu}-
{1\over2}g_{\mu\nu}\Phi^{,\sigma}\Phi_{,\sigma}],\\
&&T^\ma_{\mu\nu}=\rho e^\Phi u_\mu u_\nu.\label{matterenmo}
\end{eqnarray}
Note that the field equation (\ref{fe}) can be rewritten 
\begin{equation}\label{fe2}
\Box\Phi=-4\pi T^\ma,
\end{equation}
where $T^\ma=g^{\mu\nu}T^\ma_{\mu\nu}$. 

Matter conservation is expressed by the condition
\begin{equation}\label{matterconserv}
\nabla\cdot\vec{J}=0,  
\end{equation}
where the components of the matter current density are
$J^0=\gamma\rho$ and $J^i=\gamma\rho v^i$.  Here $v^i$ is the usual
three-velocity.  

The above equations can be generalized to a swarm of particles by
letting
\begin{equation}
m \rightarrow \sum_A m_A,~~~~u^\mu \rightarrow u^\mu_A,~~~~\mbox{etc.}
\end{equation}
(see equation (2.16) and related paragraph in ST).

Integrating (\ref{matterconserv}) over all space yields a conserved
rest mass
\begin{equation} \label{restmass}
M_0=\int J^0d^3x=\int\gamma\rho d^3x=\int\rho_0d^3x=\hbox{const}.
\end{equation}
Combining the integrals of Eq.~(\ref{enmoconserv}) and
Eq.~(\ref{matterconserv}) gives rise to a conserved total energy
at any time $t$ inside a sphere of arbitrary radius $r$ 
centered at the origin:
\begin{equation}
E_{\rm tot} = E_1 + E_2 + E_3. \label{e_tot}
\end{equation}
Here $E_1$ is the energy of the gravitational field, including a 
dynamical component,
\begin{equation}
E_1={1\over8\pi}\int^r_0{r^\prime}^2dr^\prime
\int(\Phi^2_{,t}+(\nabla\Phi)^2)d\Omega,
\end{equation}
$E_2$ is the particle's kinetic and gravitational binding energy,
\begin{equation}
E_2=\int^r_0{r^\prime}^2dr^\prime\int\rho_0(e^\Phi\gamma-1)d\Omega,
\end{equation}
and $E_3$ is the total outgoing flux of particles and radiation across
$r$, integrated over all time,
\begin{equation}
E_3=-r^2\int^t_0dt^\prime\int d\Omega
\left[{1\over4\pi}\Phi_{,t}\Phi_{,r}-\rho_0v_r(e^\Phi\gamma-1)\right].
\end{equation}
Since the total energy is conserved, and since $E_3$ vanishes initially,
$E_{\rm tot}$ at all times has to be equal to the sum of $E_1$ and $E_2$
at $t=0$,
\begin{eqnarray}
E_{\rm tot}(0) = & 
\int^r_0{r^\prime}^2dr^\prime\int d\Omega
& \big[{1\over8\pi}(\Phi^2_{,t}+(\nabla\Phi)^2)\nonumber \\
& & \left.+\rho_0(e^\Phi\gamma-1)\big] \right|_{t=0}.
\label{intmatenmoconserv}
\end{eqnarray}
As shown in ST, the total conserved mass-energy $M=\int T^{00}d^3x$
is related to $E_{\rm tot}$  according to $M=M_0+E_{\rm tot}$.

The radiative energy flux in the wave zone is
\begin{equation}
T^{0r}_\gr=-{1\over 4\pi}\Phi_{,t}\Phi_{,r}
\approx {1\over 4\pi}\Phi_{,t}^2,
\end{equation}
and the total rate of energy emission
\begin{equation}\label{energyemission}
{dE_\gr\over dt}=4\pi r^2T^{0r}_\gr \approx(r\Phi_{,t})^2.
\end{equation}
Note that these expressions become exact as $r \rightarrow \infty$.

In the weak-field, slow-motion limit, the radiation field can also be
expressed as a multipole expansion, which we will use to compare with
our QE approximation.  Since the theory involves a scalar
field, the lowest-order contribution to the radiation arises from the
monopole term.  Using the usual Green's function for a wave equation,
we follow ST and transform the wave Eq.~(\ref{fe}) into the integral form
\begin{equation}
\Phi(t,r) = -\int d^3 x' {[e^\Phi\rho]_{\rm ret}\over|{\bf x}
-{\bf x'}|} 
\approx -{1\over r}\int d^3x' \left[{e^{2\Phi}\over\tilde{u}^0}
\rho_0\right]_{\rm ret},\label{waveform}
\end{equation}
where have replaced $|{\bf x} -{\bf x'}|$ with $r = |{\bf x}|$ for
large separations, and where ``ret'' means evaluate at retarded time
$t'=t-|{\bm x}-{\bm x}'|$.  For a spherically symmetric density
distribution, the leading-order radiation field of $\Phi$ gives rise
to the monopole formula
\begin{equation}
\Phi(t,r) = - \displaystyle {4 \pi \over r}\int dr' r^{\prime2}  
	\left[\rho_0(\Phi-{1\over2}v^2)
 	+ {1\over6}r^{\prime2}\rho_{,tt}\right]_{t-r}.\label{monopole}
\end{equation}
This equation is the analogue of the ``quadrupole formula'' in general
relativity (see also the discussion below eq.~(3.8) in ST).  Note
again that scalar gravity admits gravitational radiation even in
spherical symmetry, in contrast with GR.


\subsection{Spherical matter shell}
\label{sms}

We now consider a thin, spherical shell of collisionless
particles, all of the same rest mass $m_{\rm A}$.  
At every point on the shell the particles move isotropically 
in the plane perpendicular to the radius.
In an oscillating shell, 
each particle moves about the center in a bound orbit.  
In the Newtonian limit, each orbit is a closed ellipse.

In spherical symmetry, the geodesic equation~(\ref{pe}) 
for a particle in the shell becomes
\begin{eqnarray}
{dR\over dt} & = & {\tilde{u}_r\over\tilde{u}^0},\label{r_t}\\
{d\tilde{u}_r\over dt} & = & {\tilde{u}^2_\phi\over\tilde{u}^0R^3}
-e^{2\Phi}{\Phi_{,r}\over\tilde{u}^0},\label{radiuseqn}\\
\tilde{u}_\phi & = & \hbox{const},\label{ang_t}
\end{eqnarray}
where 
\begin{equation} \label{u0}
\tilde{u}^0 = \sqrt{e^{2\Phi}+\tilde{u}_r^2+\tilde{u}^2_\phi/R^2},
\label{unitenergy}
\end{equation}
and where we have defined
\begin{equation}
\tilde{u}^a \equiv e^\Phi u^a.
\end{equation}
Each particle orbits in a plane,
conserving its orbital angular momentum $\tilde{u}_\phi$.  
Note that it
is sufficient to integrate the geodesic equations for one particle,
which then represents the entire swarm.  Note also that for a static
gravitational field, the particle energy $\tilde{u}^0$ is constant.

The particle mass density is 
\begin{equation}\label{pmd}
\rho = \sum_A {m_A\over\gamma}\delta(r - R)\delta(\theta-\theta_A)
       \delta(\phi-\phi_A)\frac{1}{r^2 \sin \theta},
\end{equation}
where $(\theta_A,\phi_A)$ are distributed isotropically on a sphere.
Inserting equation~(\ref{pmd}) into equation~(\ref{restmass}) yields 
\begin{equation}
M_0 = \sum_A m_A,
\end{equation}
so that smoothing out the particle distribution in the angular
direction, we may rewrite the density as a purely radial function,
\begin{equation}\label{mrdensity}
\rho = {M_0\over4\pi R^2\gamma} \delta(r-R),
\end{equation}
where $R$ is the radius of the shell.

In spherical symmetry, the field equation~(\ref{fe}) can be written
as
\begin{equation} \label{fess}
- \Phi_{,tt} + \frac{1}{r^2}(r^2 \Phi_{,r})_{,r} = 4 \pi e^\Phi\rho.
\end{equation}
Regularity at the origin requires the boundary condition
\begin{equation} \label{bco}
\Phi_{,r}=0,\quad\hbox{\rm at}\quad r=0,
\end{equation}
and we impose an outgoing wave boundary condition at the outer boundary,
\begin{equation} \label{bca}
(r\Phi)_{,t}+(r\Phi)_{,r}=0.
\end{equation}
The delta function
on the right hand side of~(\ref{fess}) introduces a discontinuity in
the first space derivative of $\Phi$.  Integrating the field equation
across the shell yields the jump condition
\begin{equation} \label{nonsmooth}
\Phi_{,r}|_+ - \Phi_{,r}|_- = \frac{M_0}{R^2} \, \frac{e^{2\Phi}}{\tilde{u}^0}.
\end{equation}
Note that $\Phi$ itself is continuous across the shell.  In
equation~(\ref{radiuseqn}), the force term $\Phi_{,r}$ then has to be
replaced by
\begin{equation}\label{boundiscon}
\Phi_{,r} \rightarrow {1\over2} (\Phi_{,r}|_+ +\Phi_{,r}|_-).
\end{equation}
This expression can be found by properly averaging $\Phi_{,r}$ over
an extended shell, 
and then taking the limit as the shell thickness goes to zero.


\subsection{Static solutions}
\label{stat_sol}

For static solutions, in which all particles are in circular orbits,
the field equation~(\ref{fess}) reduces to
\begin{equation} \label{staticfield}
\nabla^2\Phi= \frac{1}{r^2}(r^2 \Phi_{,r})_{,r} = 4 \pi e^\Phi\rho.
\end{equation}
In vacuum, $\Phi$ is either constant of is proportional to $r^{-1}$.
Given the boundary conditions~(\ref{bco}) and~(\ref{bca}),
we therefore find solutions of the form
\begin{equation} \label{stnypot}
 \begin{array}{rcll}
  \Phi(r)&=&-C,&\quad r\le R_{\rm S},\\
  \Phi(r)&=&\displaystyle-\frac{M_C}{r} ,&\quad r \ge R_{\rm S},
 \end{array}
\end{equation}
where $C$ and $M_C$ are constants, and where $R_{\rm S}$ is the static
equilibrium radius.  Note that $M_C$ determines the motion of distant
particles and gives rise to Kepler's laws, so that it can be
identified with a ``Coulomb'' mass, as discussed in ST.
Unlike in GR, $M_C$ does not agree with the total conserved mass-energy 
$M=M_0+E_{\rm tot}$ defined in~(\ref{restmass}) and~(\ref{e_tot}).

Since $\Phi$ is continuous across the shell, we have
\begin{equation}\label{intpot}
C = \frac{M_C}{R_{\rm S}},
\end{equation}
and using the jump condition~(\ref{nonsmooth}) yields
\begin{equation} \label{C1}
C = \frac{M_0}{R_{\rm S}} \, \frac{e^{-2C}}{\tilde{u}^0}.
\end{equation}
For a circular orbits we have $\tilde{u}^r = d\tilde{u}^r/dt = 0$,
so that equation~(\ref{radiuseqn}) yields
\begin{equation} \label{u_phi_1}
\tilde{u}^2_{\phi} = \frac{e^{-2C}}{2} R_{\rm S}^2 C,
\end{equation}
and therefore
\begin{equation} 
\tilde{u}^0 = e^{-C} (1 + C/2)^{1/2}.
\end{equation}
Inserting this into~(\ref{C1}) yields
\begin{equation} \label{C2}
C = \frac{M_C}{R_{\rm S}} = \frac{M_0}{R_{\rm S}} \, \frac{e^{-C}}{(1
+ C/2)^{1/2}}.
\end{equation}
Given $M_0/R_{\rm S}$, equation~(\ref{C2}) can be solved numerically for $C$
or, equivalently, $M_C/R_{\rm S}$.  In the Newtonian limit, where
$M_0/R_{\rm S}$ is small, we find
\begin{equation}
M_C = M_0 \, \left( 1 - \frac{5}{4} \, \frac{M_0}{R_{\rm S}} \right).
\end{equation}
In the limit of large $M_0/R_{\rm S}$, $C$ and $M_C/R_{\rm S}$ scale
with $\log(M_0/R_{\rm S})$, implying that unlike in GR, there is no
maximum compaction $M_C/R_{\rm S}$.

We can also combine equations~(\ref{u_phi_1}) and~(\ref{C2}) to obtain
\begin{equation} \label{C3}
\frac{e^{-2C}}{(2C - C^2)^{1/2}} = \frac{\tilde{u}_{\phi}}{M_0}.
\end{equation}
Given $M_0$ and the angular momentum $\tilde{u}_{\phi}$, this equation
can be solved for $C$.  The result can then be inserted
into~(\ref{u_phi_1}) to yield the radius $R_{\rm S}$. 

The oscillations which we consider in this paper are damped, and ultimately
give rise to a static equilibrium state.  During the damping process, both the 
rest mass $M_0$ and the angular momentum $\tilde{u}_{\phi}$ are conserved.
Given any initial non-equilibrium configuration 
with $M_0$ and $\tilde{u}_{\phi}$,
we can therefore use~(\ref{C3}) to determine 
the final equilibrium configuration to which
the oscillating shell will ultimately settle down.


\subsection{Newtonian limit}
\label{newton_limit}

In the Newtonian limit, i.e.~in the limit of weak fields and slow
velocities, an analytic solution for a periodically oscillating shell
can be derived (see Sect.~VI of ST).  Consider a static shell of rest
mass $M_0$ and radius $R_i$, and reduce all velocities instantaneously
by a factor $\xi$ at $t=0$.  The individual particles comprising the shell 
then all move in
elliptical orbits with the same period, eccentricity and semimajor axes
satisfying
\begin{equation}\label{varad}
R=R_ix(t),
\end{equation}
where $x(t)$ is given by the usual parametric equations for an
elliptic orbit:
\begin{equation}\label{elliptic}
 \begin{array}{l}
  x=a(1+e\cos u),\\[3mm]
  t=\displaystyle{P\over2\pi}(u+e\sin u),
 \end{array}
\end{equation}
Here the semimajor axis, eccentricity and period are, respectively 
\begin{equation}
 \begin{array}{l}
  a=\displaystyle{1\over2-\xi^2},\\[3mm]
  e=1-\xi^2,\\[3mm]
  \displaystyle{P=2\pi \left(2R_i^3\over M_0(2-\xi^2)^3\right)^{1/2}}.
 \end{array}
\end{equation}
The radial and tangential particle velocities are given by
\begin{equation}
\begin{array}{l}
v_r = \displaystyle \frac{\dot x}{x}R, \\[3mm]
v_{\phi} = \displaystyle \xi \frac{r}{x^2} \left(\frac{M_0}{R_i^3} 
\right)^{1/2}.
\end{array}
\end{equation} 
We have used this analytic solution extensively to test our code in
the Newtonian limit.

Inserting the analytic solution into Eq.~(\ref{monopole}) and
differentiating with respect to time yields the wave amplitude
\begin{equation}\label{waveamp}
\Lambda=r\lambda=-{4\over3}{M_0^2\over R_i}
\left[{\dot x\over x^2}\right]_{t-r}\mbox{(wave zone)},
\end{equation}
The total rate of energy emission can now be found from
Eq.~(\ref{energyemission}),
\begin{equation}\label{monoerad}
{dE\over dt}={16\over9}{M_0^4\over R^2_i}
\left[{\dot x^2\over x^4}\right]_{t-r}.
\end{equation}


\section{Quasi-equilibrium scheme}
\label{QES}

In the QE approximation, we assume that the orbital decay time is much
longer than the orbital period.  On an orbital timescale, the effect
of the gravitational radiation can then be neglected, and it is
reasonable to assume that each orbit is determined by the ``Coulomb''
part of the gravitational field rather than the radiative part.
This suggests that we can neglect the second time derivative in the
field equation~(\ref{fe}), so that the field then satisfies the elliptic
equation
\begin{equation} 
\nabla^2 \Phi =  4 \pi e^\Phi \rho_{\rm QE}.
\end{equation}
Similarly, QE approximations in GR typically lead to elliptic
equations for the gravitational field components (see,
e.g.,~\cite{wm95,bcsst,irrotational}).  This approximation greatly
simplies the problem, since the gravitational field no longer has any
dynamical degrees of freedom.

Since no radiation is generated in the QE approximation, the system is
strictly conservative and there is a conserved energy $E_{\rm QE}$.
This energy can be derived by multiplying equation~(\ref{radiuseqn})
with $2\tilde{u}_r$, which yields
\begin{equation} \label{eqe_1}
{d\tilde{u}_r^2\over dt}=-{d\over dt}{\tilde{u}_\phi^2\over
R^2}-2e^{2\Phi_{\rm sh}} \Phi_{,r} {dR\over dt}.
\end{equation}
Here $\Phi_{\rm sh} = \Phi(R)$, and we have used conservation of
angular momentum $\tilde{u}_\phi = const$.  From
Eqs.~(\ref{boundiscon}) and~(\ref{stnypot}) we also have
\begin{equation}
\Phi_{,r} = -{\Phi_{\rm sh}\over2R}.
\end{equation}
Using~(\ref{C1}), we can now rewrite eq.~(\ref{eqe_1}) as
\begin{equation}
{d\over dt}(\tilde{u}_r^2+{\tilde{u}_\phi^2\over
R^2})=-{\tilde{u}^0\Phi_{\rm sh}^2\over M_0}{dR\over dt}.
\end{equation}
Expressing the left hand side in terms of $\tilde{u}^0$ (eq.~(\ref{u0}),
we finally find
\begin{equation} \label{toteng}
E_{\rm QE} \equiv M_0\tilde{u}^0+{1\over2}R\Phi_{\rm sh}^2=const.
\end{equation}

In order to relate $E_{\rm QE}$ to the conserved total energy $E_{\rm tot}$ 
(eq.~(\ref{e_tot})), we evaluate the latter in the QE approximation
by inserting the solution~(\ref{stnypot}) and by setting the radiative
components of the field equations to zero: $\Phi_{,tt} = \Phi_{,t} = 0$.
We then find
\begin{eqnarray}
E_1&=&{1\over2}\int^r_0(\nabla\Phi)^2{r'}^2dr',\nonumber\\
&=&-{1\over2}\Phi_{\rm sh}^2R^2\int^r_R{dr'\over{r'}^2}
=-{1\over2}\Phi_{\rm sh}^2R^2({1\over r}-{1\over R})
\end{eqnarray}
and
\begin{equation}
E_2 = \int^r_0{r'}^2dr'\int\rho_0(\tilde{u}^0-1)d\Omega=M_0(\tilde{u}^0-1).
\end{equation}
The radiative contribution $E_3$ in~(\ref{e_tot}) vanishes identically.
We therefore have
\begin{equation}
E_{\rm tot} = E_1 + E_2 =
M_0(\tilde{u}^0-1)-{1\over2}\Phi_{\rm sh}^2R^2({1\over r}-{1\over R}),
\end{equation}
and find for $r\rightarrow\infty$
\begin{equation}\label{e1e2}
E_{\rm tot} = E_{\rm QE} - M_0 = const.
\end{equation}
Note that this energy is conserved only when $E_1$ is evaluated with
$r\rightarrow\infty$, so that it includes the entire potential energy
of the longitudinal (or Coulomb-like) gravitational fields.

In the Newtonian limit,
\begin{equation}
\tilde{u}^0 = \gamma e^{\Phi_{\rm sh}} = {e^{\Phi_{\rm sh}}\over\sqrt{1-v^2}}
\approx 1+\Phi_{\rm sh}+{1\over2}v^2.
\end{equation}
Since to lowest order $\Phi_{\rm sh} \approx -M_0/R$, we have
\begin{equation}
E_{\rm tot} = {1\over2}M_0v^2-{M_0^2\over2R}
\end{equation}
in the Newtonian limit.


\begin{figure}[tb]
\begin{center}
\leavevmode
\epsfxsize=3.3in
\epsfysize=4.2in
\epsffile{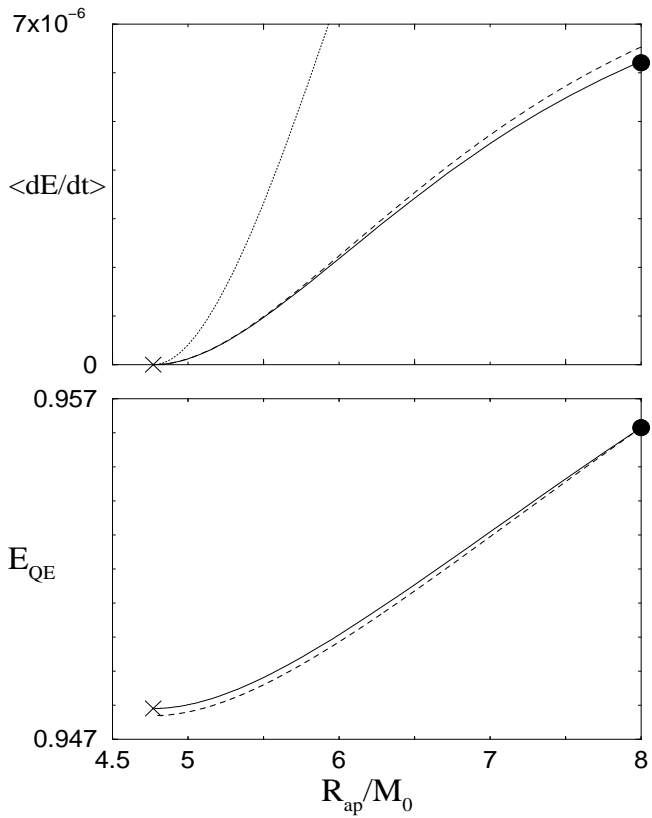}
\end{center}
\caption{ The gravitational wave luminosity (upper panel) and the
total energy (lower panel) as functions of apocenter radius (for a
moderately strong field configuration, $R_i = 8 M_0$, see
Sec.~\ref{sec4b} below).  The dotted line denotes the result from the
monopole formula (\ref{monoerad}), the dashed line shows the QE
result, and the solid line shows the integration of the exact
equations.  In the lower panel, the dashed line is computed
from eq.~(\ref{toteng}), and the solid line is computed from 
$E_{\rm QE}\rightarrow E_1 + E_2 + M_0$ as in eq.~(\ref{e_tot}) with
$r \rightarrow \infty$.}
\label{fig1}
\end{figure}

In the bottom panel of Fig.~\ref{fig1}, we show $E_{QE}$ as a function
of apocenter radius $R_{\rm ap}$ for constant angular momentum
$\tilde{u}_{\phi} = 1.17M_0^2$ (solid
line).  As we show in Appendix~\ref{mine}, the
turning point of this curve corresponds to an equilibrium
configuration in which all particles move in circular orbits (see also
Sec.~\ref{stat_sol}).  As the damped oscillations of the matter shell
radiates energy, the shell ``slides down'' the energy curve in
Fig.~\ref{fig1}, until it settles down to static equilibrium
at the curve's minimum.  We also include in the bottom
panel of Fig.~\ref{fig1} the result of an exact evolution (dashed
line), in which we started the oscillation with a static shell model at $R_i
= 8 M_0$, and then suddenly reduced all particle velocities 
by a factor of $\xi = 0.7$,
so that the particle's angular momentum
is again $\tilde{u}_{\phi} = 1.17M_0^2$.

We now adopt a perturbative approach, in which we insert the
predetermined QE matter density $\rho_{\rm QE}$ as a source to the
fully dynamical field equations
\begin{equation}\label{qsfe}
\Box\Phi(r,t)= 4\pi e^\Phi\rho_{\rm QE},
\end{equation}
where $\rho_{\rm QE}$ is given by eqn.~(\ref{mrdensity}) 
using the QE solution for $R(t)$
(compare with the ``hydro-without-hydro'' approach as
suggested in~\cite{bhs}).  Integrating this equation then yields the
periodic gravitational wave form and luminosity $dE/dt$ for each QE
configuration of a certain apocenter radius.   In practice, we determine
$dE/dt$ by averaging over one period $P$
\begin{equation} \label{de}
<\frac{dE}{dt}> = \frac{1}{P} \int_0^P \frac{dE}{dt} dt'.
\end{equation}
We show $<dE/dt>$ as computed from QE models in the top panel of
Fig.~\ref{fig1} (solid line) and compare both with results from the
integration of the exact equations (dashed line) and the monopole
formulae (dotted line).  From this plot it is already obvious that the
QE ``hydro-without-hydro'' approach provides a much better
approximation than the monopole formula.  Combining $<dE/dt>$ for
several values of $R_{\rm ap}$ with the derivative of the QE energy
$dE_{\rm QE}/dR_{\rm ap}$ then yields the damping rate
\begin{equation}\label{chain}
{dR_{\rm ap} \over dt} = 
{<dE/dt> \over dE_{\rm QE}/dR_{\rm ap}}.
\end{equation}
(This is analogous to eq.~(1) in~\cite{dbs} 
for binary neutron star inspiral.)
Note that at the final
equilibrium radius $R = R_{\rm S}$, the numerator $<dE/dt>$ vanishes more
rapidly then the denominator $dE_{\rm QE}/dR_{\rm ap}$, so that
$dR_{\rm ap}/dt$ smoothly approaches zero.  

The complete wavetrain of the damped oscillations can be assembled by
suitably parameterizing the wave signals for different values of $R_{\rm
ap}$ (compare with eq.~(2) in~\cite{dbs}).  We provide details of this
parameterization in Appendix~\ref{proc}.


\section{Numerical results}
\label{numres}

In this section, we compare results from the QE approach, the exact
integration and the monopole formula for weak, moderately strong and 
ultra-strong fields.  
For each case, we prepare a static shell solution as in
Sec.~\ref{stat_sol} for a certain value of $M_0/R_i$, where $R_i$ is
the initial apocenter radius, and then reduce all velocities and hence
angular momenta by a factor of $\xi = 0.7$.  The particle orbits are
then out of equilibrium and start a damped oscillation, until they
lose sufficient energy by gravitational radiation to
settle down into the final static equilibrium corresponding to 
the reduced value of the angular momentum.

Given the above scenario, the appropriate initial
conditions for the gravitational field are
\begin{equation}
\Phi=\Phi_{\rm static},\qquad\Phi_{,t}=0,
\end{equation}
where $\Phi_{\rm static}$ is given by Eqs.~(\ref{stnypot}) and~(\ref{C3}).
The particle initial data are
\begin{equation}
R=R_{\rm S},\qquad\tilde{u}_r=0,\qquad\mbox{and }\tilde{u}_\phi=
\xi\tilde{u}_{\phi}^{\rm static},
\end{equation}
where $\tilde{u}_{\phi}^{\rm static}$ is given 
by~(\ref{intpot}) and~(\ref{u_phi_1}).
Thus the (non-equilibrium) shell begins at rest 
with all the particles at their apocenter positions.

For numerical reasons (involving the regridding algorithm as described in
Appendix~\ref{numet}), we found it convenient to impose outer boundary
conditions at $5 R_i$ (except for the strong field case, where we
choose $50 R_i$ for the outer boundary).  We resolve the interior region of
the shell with 50 gridpoints, and the exterior region with 200 gridpoints.
The field integration scheme is adopted from ST and summarized in 
Appendix~\ref{numet}.

\subsection{Weak field configuration ($R_i=1000M_0$)} 

\begin{figure}[tb]
\begin{center}
\leavevmode
\epsfxsize=3.3in
\epsfysize=2.2in
\epsffile{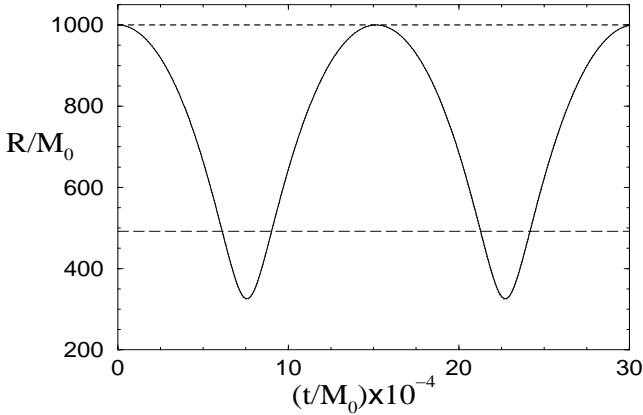}
\end{center}
\caption{Evolution of the radius as a function of time for a matter
shell with the initial radius $R_i=1000M_0$ and velocity cut-down 
factor $\xi=0.7$.  The solid
line is the result from the integration of the exact equations.  The
dotted line, which completely coincides with the solid line, marks the
analytic Newtonian result of Sec.~\ref{newton_limit}.  
The short-dashed line is
the envelope of apocenter radii oscillations according to the QE
approach.  The long-dashed line marks the final equilibrium radius of
the shell.  }
\label{fig2}
\end{figure}

We first study a weak field configuration with $R_i=1000M_0$.  After
having reduced the angular momentum be a factor of $\xi = 0.7$, this
configuration will ultimately settle down to a final radius of $R_f =
R_{\rm S} = 491.6M_0$.  In Fig.~\ref{fig2} we compare the evolution
of the shell's radius as found from the integration of the exact
equations (Sec.~\ref{sms}), the QE approach (Sec.~\ref{QES}), 
and the analytic Newtonian result (Sec.~\ref{newton_limit}).  
As expected, the agreement between
the different approaches is excellent.


\begin{figure}[tb]
\begin{center}
\leavevmode
\epsfxsize=3.3in
\epsfysize=2.2in
\epsffile{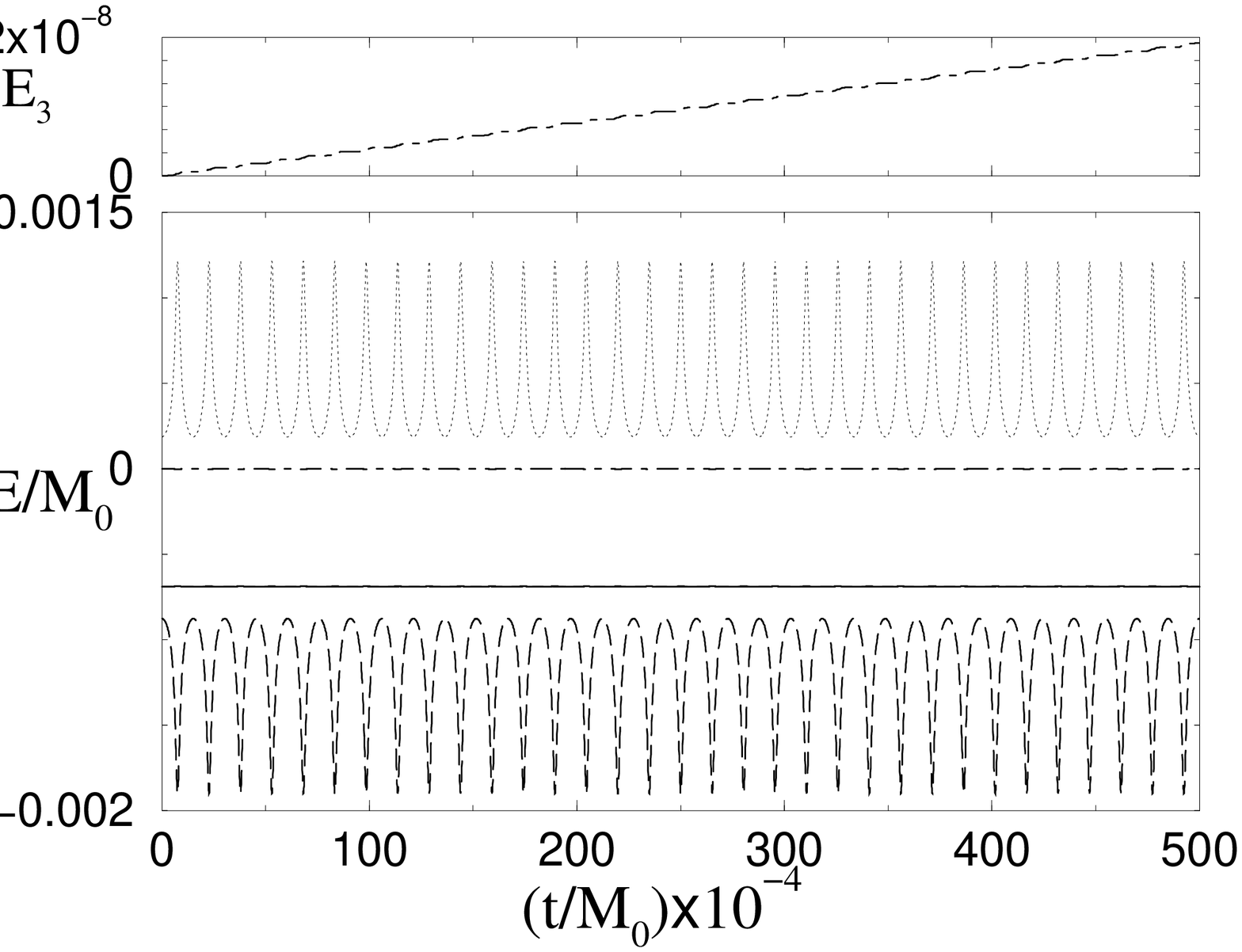}
\end{center}
\caption{
Contributions to the total energy as a function of time for
the weak field case $R_i=1000M_0$.  The solid line shows the total
energy $E_{\rm tot}$ (see eq.~(\ref{e_tot}).  The field energy $E_1$
is marked by the dotted line, the particle energy $E_2$ by the
long-dashed line, and the integrated flux $E_3$ by the dot-dashed line
(see also top panel).  We also include the initial value of $E_{\rm
tot}(0)$, marked by a horizontal dashed line.  
The agreement between the solid and
the dashed line demonstrates energy conservation and 
is a measure of the accuracy of our code.}
\label{fig3}
\end{figure}

In Fig.~\ref{fig3} we show that energy is conserved to very high
accuracy in our code.  Here, we evaluate the conserved integral
(\ref{intmatenmoconserv}) at radius $r=1600M_0$.
The plot also shows that the
radiated energy $E_3$ (integrated flux) is very small compared with the
other energy terms.

\begin{figure}[tb]
\begin{center}
\leavevmode
\epsfxsize=3.3in
\epsfysize=4.2in
\epsffile{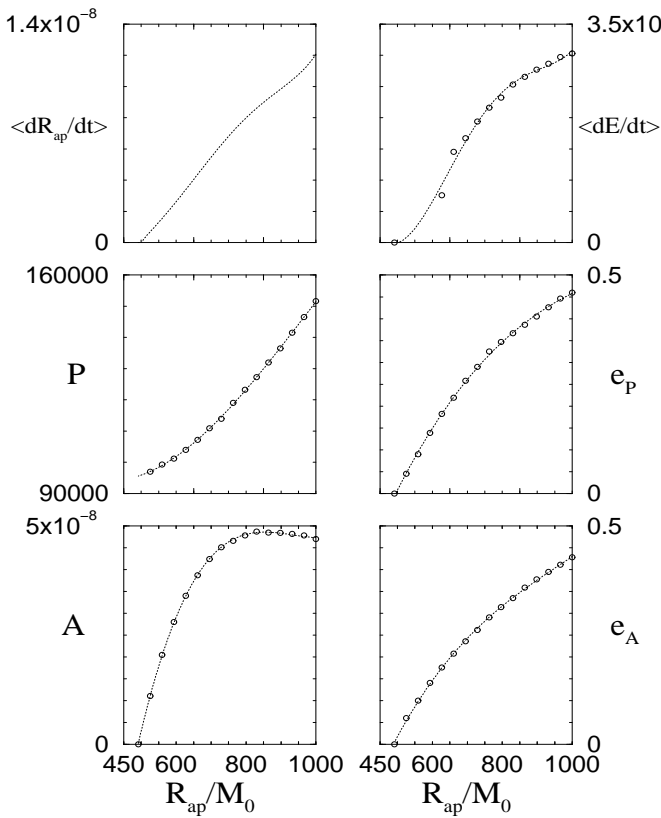}
\end{center}
\caption{Calculated QE data and their
fitting curves as functions of apocenter radius for the $R_i=1000M_0$
case.  The circles are the calculated data points, 
and the dotted lines are the fitted curves.}
\label{fig4}
\end{figure}

We show the QE parameters from which the QE waveform is constructed in
Fig.~\ref{fig4}.  The gravitational radiation luminosity $<dE_{\rm
QE}/dt>$ is computed by integrating eq.~(\ref{qsfe}) for 13 different
values of $R_{\rm ap}$ and interpolating between the results.
Using~(\ref{de}) and~(\ref{chain}), 
this can be combined with $dE_{\rm QE}/dR_{\rm ap}$
to yield the damping rate $dR_{\rm ap}/dt$.  The 4 parameters $A$,
$e_A$, $P$, and $e_P$  
of the waveform~(\ref{qetime}) and~(\ref{qewave}) at
different values of $R_{\rm ap}$ are determined by nonlinear data
fitting (Appendix~\ref{construct}).  
As expected, the values of $e_A$ and $e_P$ are very similar
in the weak field case.


\begin{figure}[tb]
\begin{center}
\leavevmode
\epsfxsize=3.3in
\epsfysize=3in
\epsffile{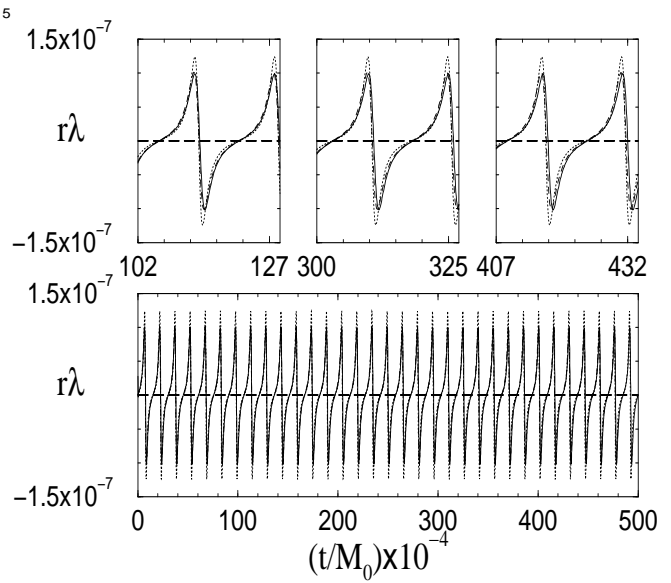}
\end{center}
\caption{Wave amplitude $\lambda$ (multiplied by $r$) as a function of
time at $r=1600M_0$ for the weak field case ($R_i=1000M_0$).  The
solid line shows the integration of the exact equations, the dashed
line marks the QE result, and the dotted line is the monopole
radiation obtained from the analytic Newtonian solution (see
Sec.~\ref{newton_limit}).}  
\label{fig5}
\end{figure}

In Fig.~\ref{fig5} we compare the waveforms as obtained from the
integration of the exact equations (solid line), the QE approach
(dashed line) and the Newtonian analytic solution (dotted line) for
the first 30 periods.  The QE approach can reproduce the exact result
very well.  Note also that the QE scheme is very efficient: given an
interpolation between the data for a small set of apocenter radii
$R_{\rm ap}$, the entire wavetrain, which in this case would take
thousands of cycles and would, in an exact integration, be dominated
by accumulation of numerical noise, can be constructed very easily.


\subsection{Moderately strong field configuration ($R_i=8M_0$)}
\label{sec4b}

We now turn to a configuration with a moderately strong field, $R_i=8M_0$.
In this case, the oscillation damps out much more quickly
due to radiation reaction, and we can
integrate the exact equations until equilibrium has essentially been reached.
This occurs at $R=4.77M_0$.

\begin{figure}[tb]
\begin{center}
\leavevmode
\epsfxsize=3.3in
\epsfysize=2.2in
\epsffile{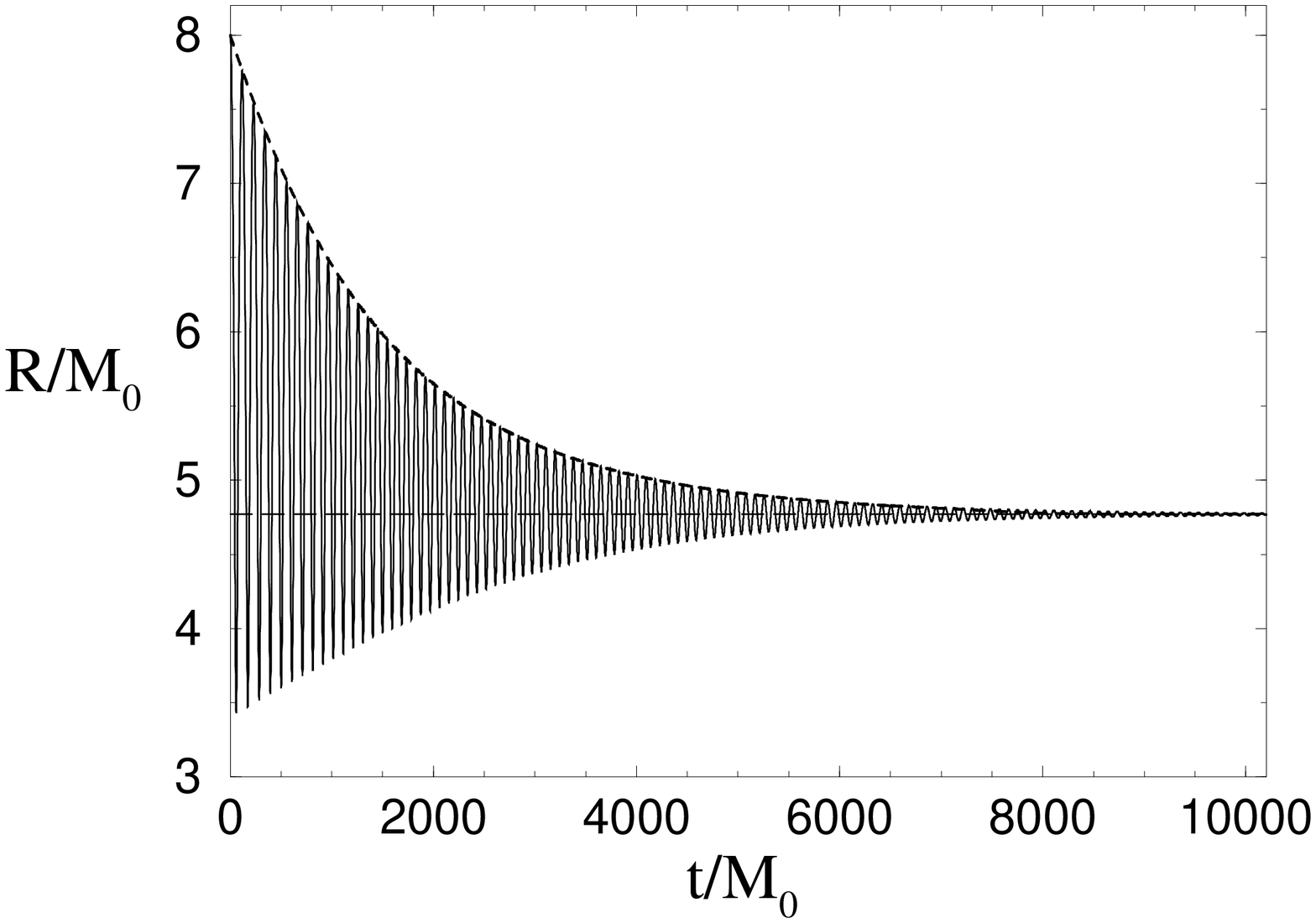}
\end{center}
\caption{Evolution of the shell radius as a function of time for a matter
shell for the moderate field case ($R_i=8M_0$). Labeling is
the same as in Fig.~\ref{fig2}, except that we do not include
the Newtonian result.}
\label{fig6}
\end{figure}

\begin{figure}[tb]
\begin{center}
\leavevmode
\epsfxsize=3.3in
\epsfysize=4.2in
\epsffile{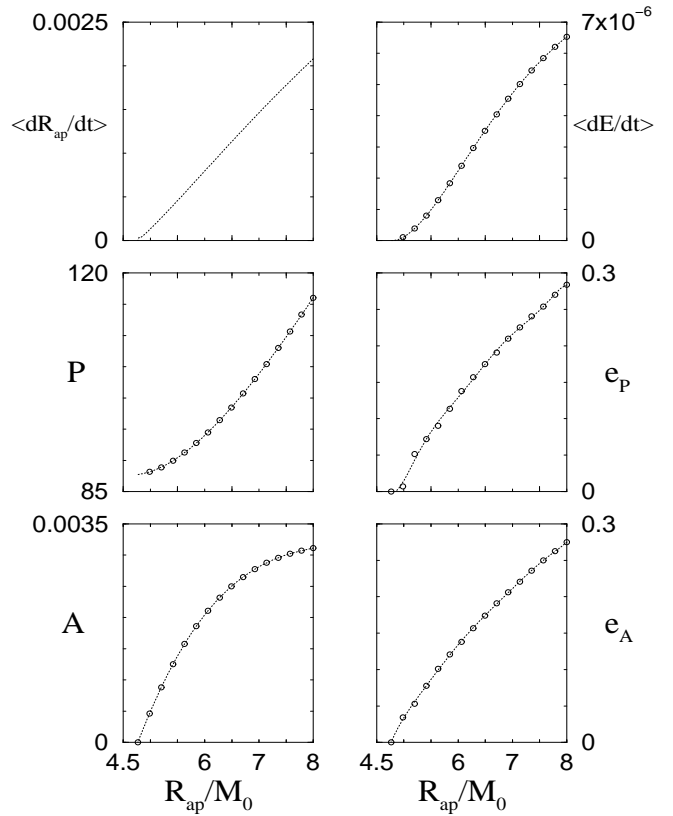}
\end{center}
\caption{Calculated data QE and their
fitting curves as functions of apocenter radius for the moderately
strong field case ($R_i=8M_0$).}
\label{fig7}
\end{figure}

\begin{figure}[tb]
\begin{center}
\leavevmode
\epsfxsize=3.3in
\epsfysize=2.2in
\epsffile{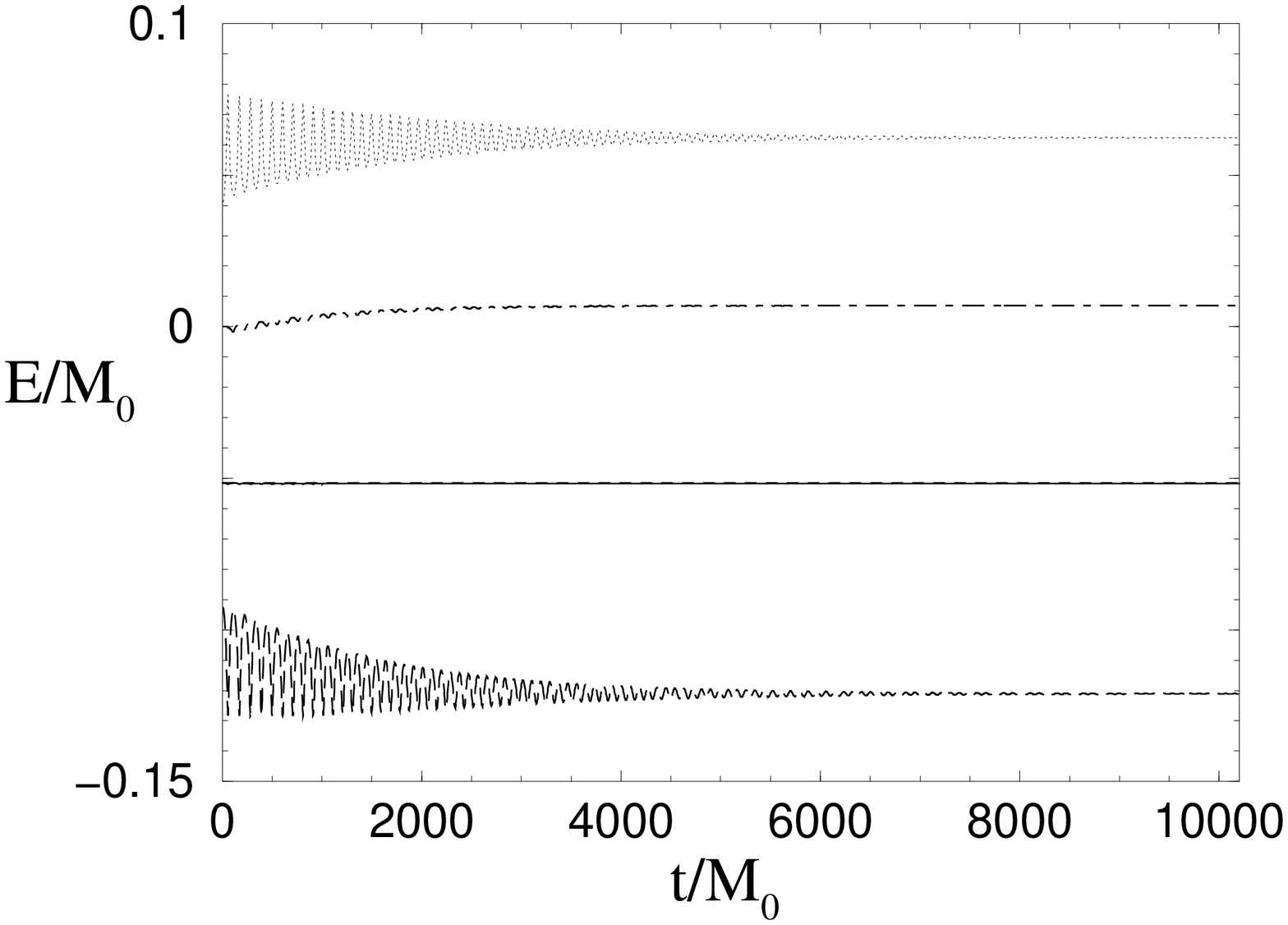}
\end{center}
\caption{Energy conservation as a function of time for
the moderately strong field case ($R_i=8M_0$). Labeling is 
the same as in Fig.~\ref{fig3}.} 
\label{fig8}
\end{figure}

We show the evolution of the shell's radius in Fig.~\ref{fig6}.  This
plot includes the result from the integration of the exact equations
(solid line) as well as the ``envelope'' $R_{\rm ap}(t)$ as found in
the QE approach.  Fig.~\ref{fig7} shows the QE parameters which we
have constructed for this configuration.  We plot the energy
contributions as a function of time in Fig.~\ref{fig8}, and find that
energy is conserved to about $0.4\%$.  Computing the exact solution
takes only a few CPU hours on, for example, an SGI O2 workstation.


\begin{figure}[tb]
\begin{center}
\leavevmode
\epsfxsize=3.3in
\epsfysize=3in
\epsffile{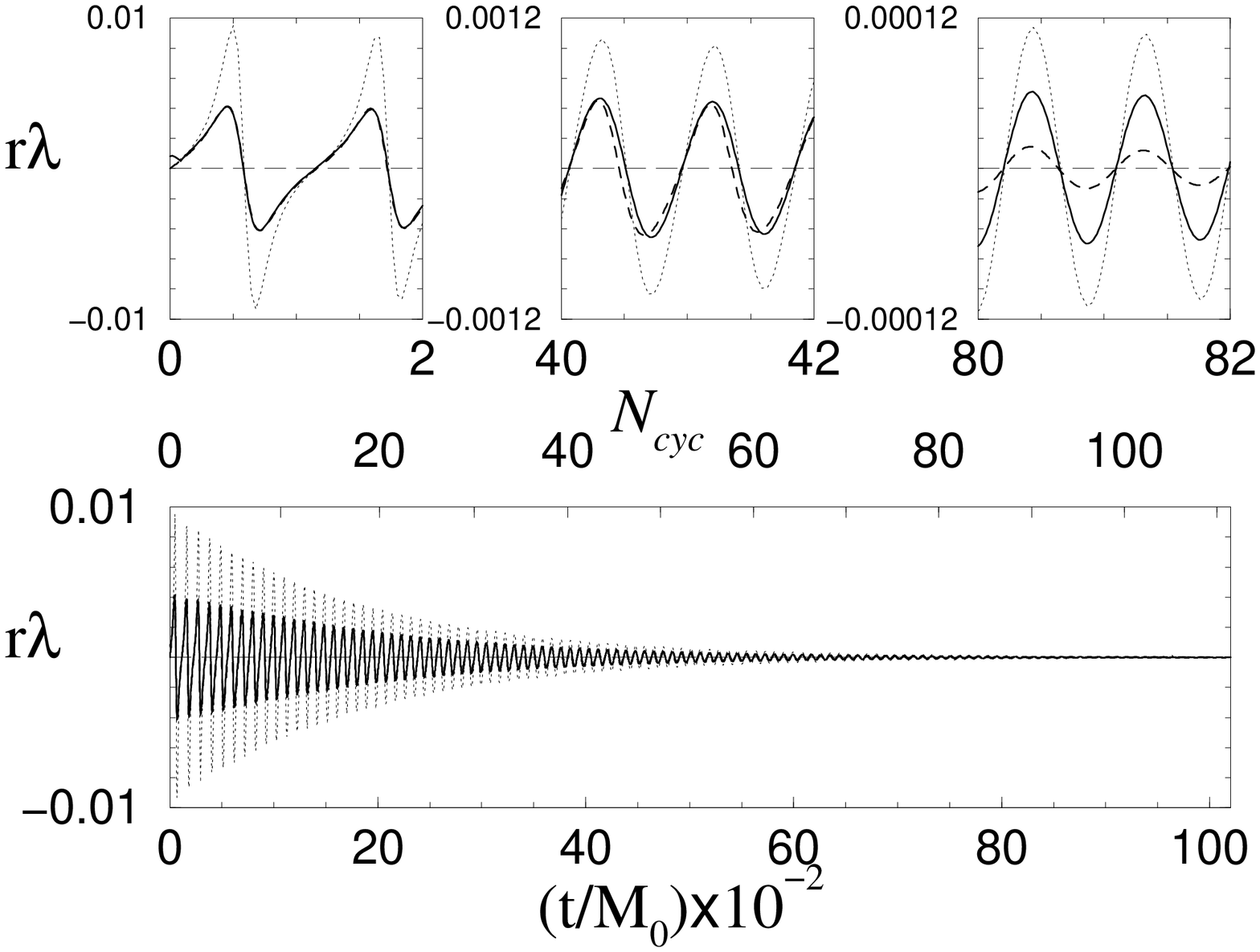}
\end{center}
\caption{Wave amplitude $\lambda$ (multiplied by $r$) as a function of
time as measured at $r=10M_0$ for the moderate field case
($R_i=8M_0$).  Labeling is the same as in Fig.~\ref{fig5}, except 
that the dotted line denotes the monopole formula applied to the
oscillating QE shell.}
\label{fig9}
\end{figure}

Fig.~\ref{fig9} shows three waveforms of the oscillating shell.
Here we compare the results of
the QE and exact integrations with the result of the monopole formula
when applied to the oscillating QE configurations.  
The QE waveform and the exact
waveform agree very well up to late times ($t \sim 7000 M_0$), at
which point the oscillation amplitude has decreased by about a factor
of 100.  At this point the integration of the exact equations has
accumulated substantial numerical noise, and may actually be less
accurate than the QE approach.  The waveform from the monopole
approach disagrees with the exact one even at very early times,
showing that the QE approach is much more reliable for moderately
strong fields.


\subsection{Ultra-Strong field configuration ($R_i=0.1M_0$)}

\begin{figure}[tb]
\begin{center}
\leavevmode
\epsfxsize=3.3in
\epsfysize=2.2in
\epsffile{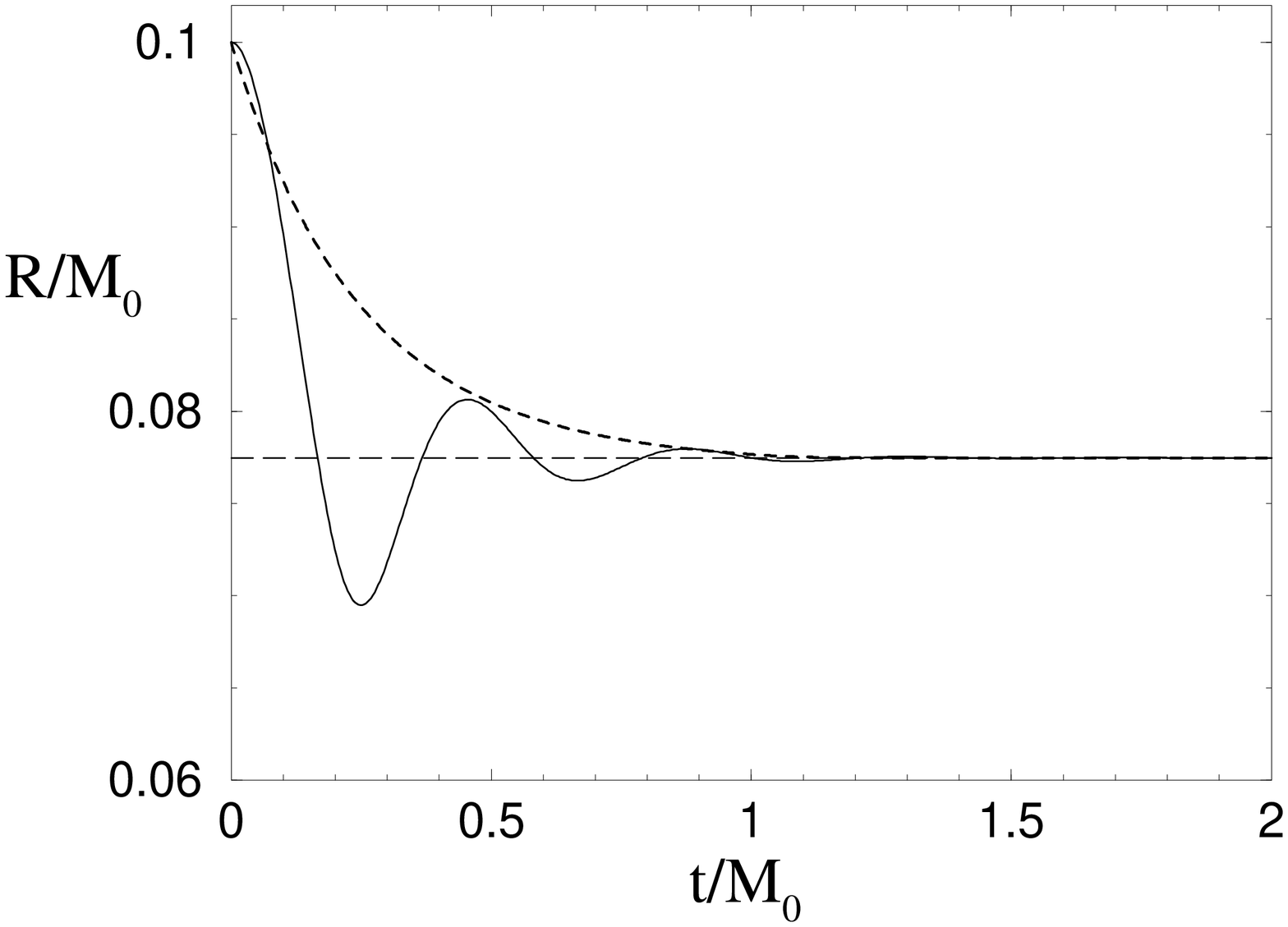}
\end{center}
\caption{Evolution of the radius as a function of time for the ultra-strong
field case ($R_i=0.1M_0$).  Labeling is the same as in
Fig.~\ref{fig2}, except that we do not include the Newtonian result.}
\label{fig10}
\end{figure}

\begin{figure}[tb]
\begin{center}
\leavevmode
\epsfxsize=3.3in
\epsfysize=4.2in
\epsffile{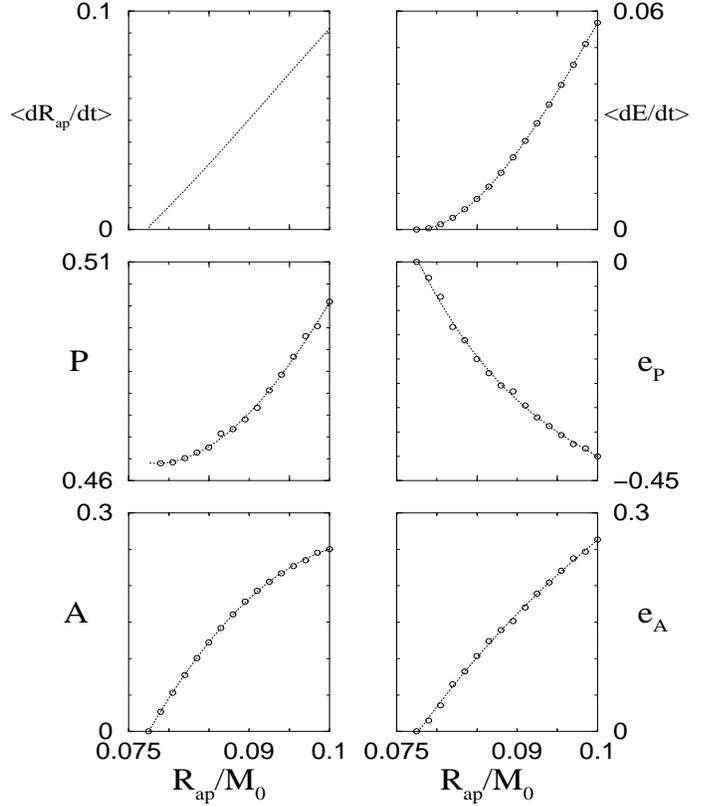}
\end{center}
\caption{Calculated QE data and their
fitting curves as functions of apocenter radius for the ultra-strong field
case ($R_i=0.1M_0$).  Labeling is the same as in Fig.~\ref{fig4}}
\label{fig11}
\end{figure}

\begin{figure}[tb]
\begin{center}
\leavevmode
\epsfxsize=3.3in
\epsfysize=2.5in
\epsffile{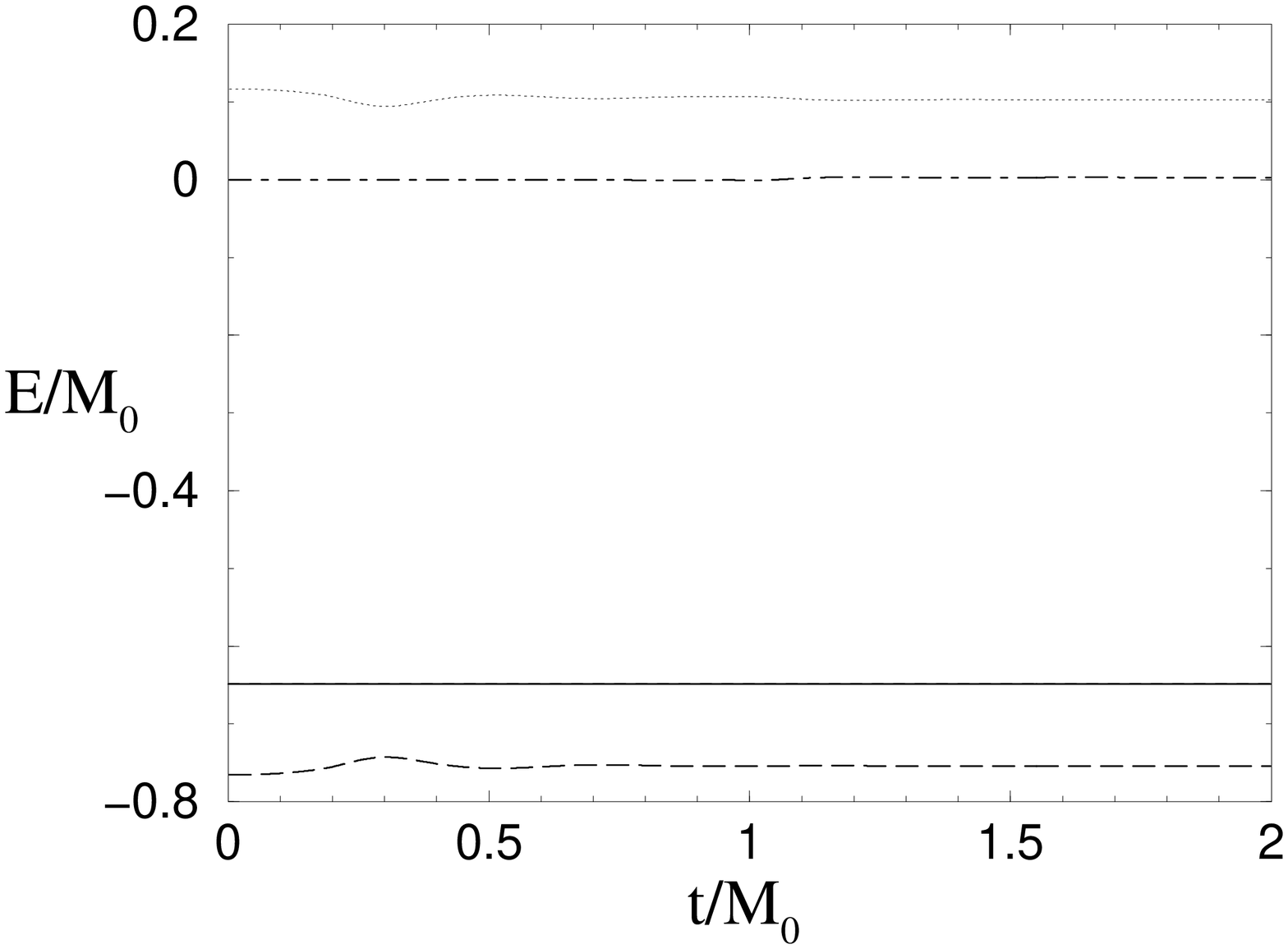}
\end{center}
\caption{Energy conservation as a function of time for
the strong field case ($R_i=0.1M_0$). Labeling is the same as in
Fig.~\ref{fig3}}
\label{fig12}
\end{figure}

For very strong fields, the oscillations are damped so strongly that
our QE assumption no longer holds, and we therefore expect our QE
approach to break down (see discussion below).  Fig.~\ref{fig10}
shows that the QE envelope $R_{\rm ap}(t)$ no longer matches the
exact result very well.  We show the fitting parameters for this case
in Fig.~\ref{fig11}, and the energy contributions in Fig.~\ref{fig12} 
which still obey energy conservation quite well.


\begin{figure}[tb]
\begin{center}
\leavevmode
\epsfxsize=3.3in
\epsfysize=2.6in
\epsffile{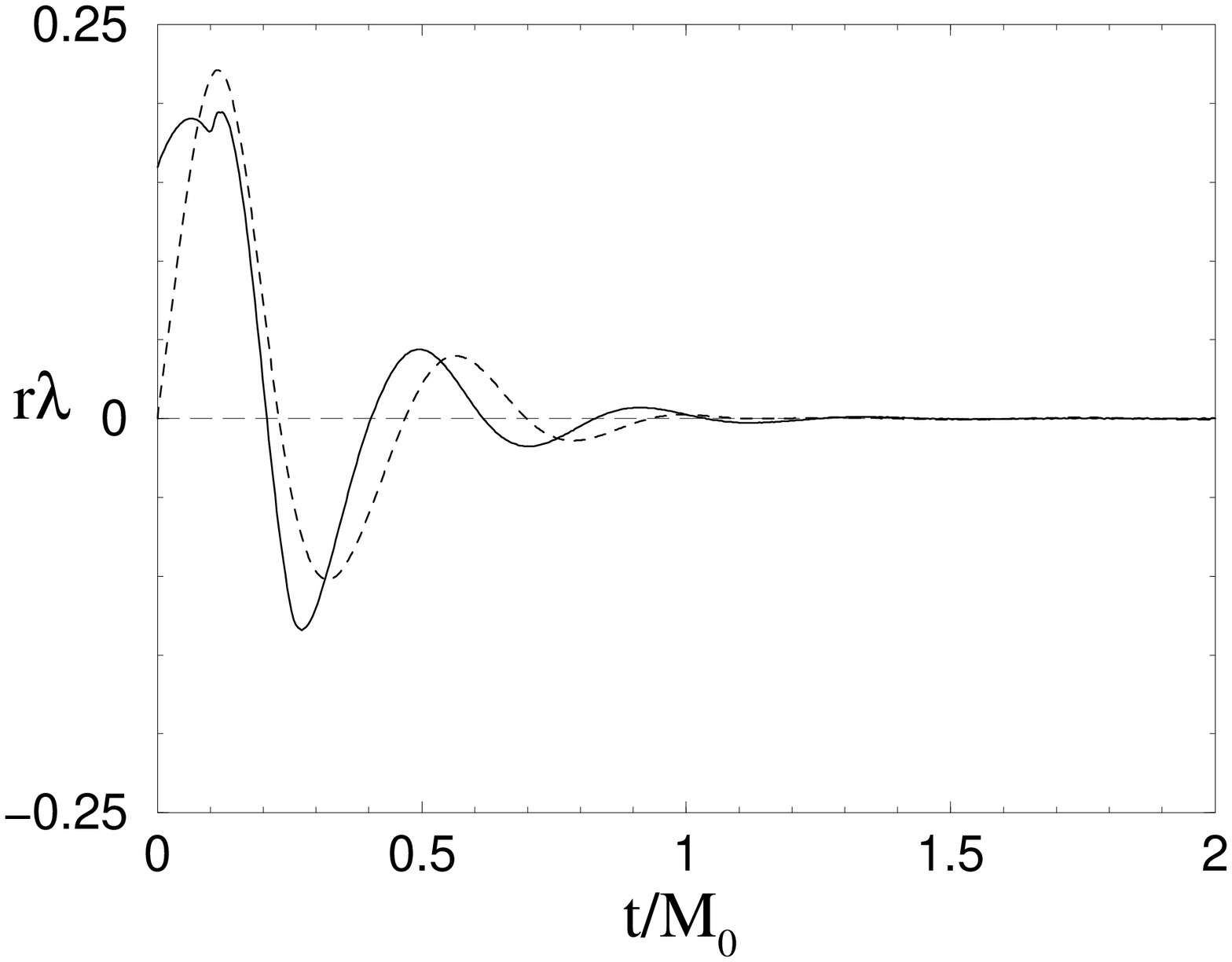}
\end{center}
\caption{Wave amplitude $\lambda$ (multiplied by $r$) as a function of
time as measured at $r=0.8M_0$ for the ultra-strong field case
($R_i=0.1M_0$).  Labeling is the same as in Fig.~\ref{fig5}, except
that we do not include the Newtonian result.  The small bump at early
times is a numerical artifact and is due to slightly imprecise initial
data.  }
\label{fig13}
\end{figure}

In Fig.~\ref{fig13} we compare the waveforms from the exact
integration and the QE approach, and find the expected disagreement.
Even here, however, the overall shape of the QE waveform is not far off.
We do not include the monopole approximation, since its predictions do
not even fit on the same scale.


We can illustrate the breakdown of the QE approach by the following 
simple argument.  Figures~\ref{fig4}, \ref{fig7} and~\ref{fig11} show
that  $dR_{\rm ap}/dt$ decays like
\begin{equation}
{dR_{\rm ap}\over dt}\approx -\alpha(R_{\rm ap}-R_f),
\end{equation}
where the value of $\alpha$ depends on $R_i/M_0$ and 
can be read off of the above
figures.  As a result, $R_{\rm ap}$ decreases exponentially,
\begin{equation}
{R_{\rm ap}-R_f\over R_i-R_f}=e^{-\alpha t}.
\end{equation}
The timescale $t_{\rm rad}=\alpha^{-1}$ can now be compared to the 
orbital timescale $t_{\rm orb}$.  Inserting numbers for the three cases, 
we find
\begin{equation}
Q\equiv{t_{\rm rad}\over t_{\rm orb}}\approx\left\{
 \begin{array}{ll}
  280000,&R_i=1000M_0,\\
  &\\
  15,&R_i=8M_0,\\
  &\\
  0.5,&R_i=0.1M_0.
 \end{array}\right.
\end{equation}
For the QE assumptions to hold we need $Q\gg1$, which obviously holds
only for  $R_i=1000M_0$ and  $R_i=8M_0$, but not for the ultra-strong field
case  $R_i=0.1M_0$.  It is therefore not surprising that the QE approach
breaks down in this case.

\begin{figure}[tb]
\begin{center}
\leavevmode
\epsfxsize=3.3in
\epsfysize=2.8in
\epsffile{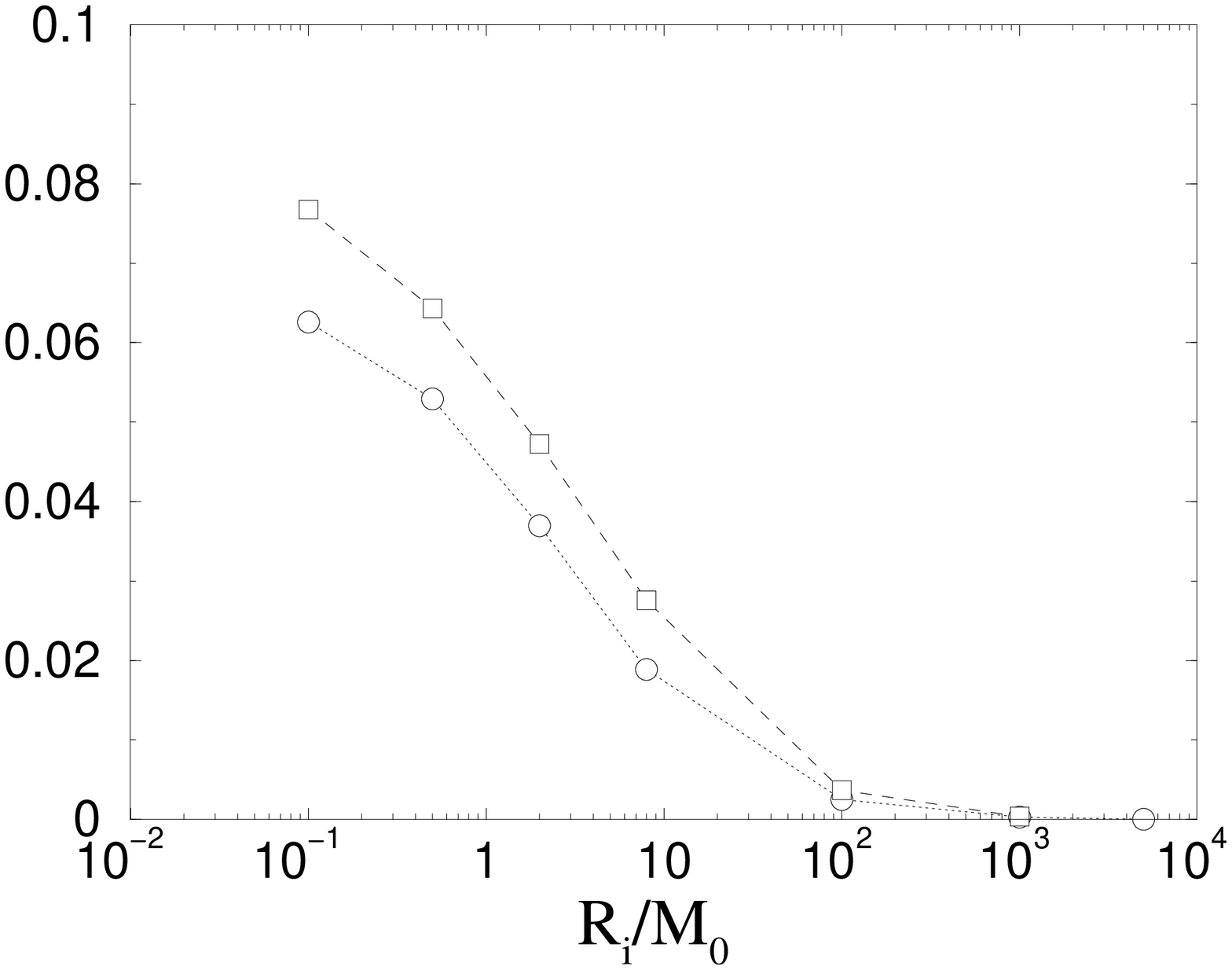}
\end{center}
\caption{The maximal value of the square of radial velocity $v_r^2$
(circles linked by the dotted line) and the dynamical part of the
field equation $R\Phi_{,tt}/2\Phi_{,r}$ (squares linked by the dashed
line) vs the initial radius.  The size of these terms is a rough
measure of the expected relative error in the QE approach.}
\label{fig14}
\end{figure}

We may also check the QE approximation, which neglects
$\Phi_{,tt}$ in the field equation~(\ref{fe}), for self-consistency.
Assuming that $\Phi \sim 1/R$, we estimate its magnitude to be
\begin{equation} 
\Phi_{,tt} \sim {\ddot R\over R^2}-{2\dot{R}^2\over R^3}
\sim \frac{\dot{R}^2}{R^3}
\end{equation}
and similarly
\begin{equation}
\nabla^2 \Phi \sim \frac{1}{R^3}.
\end{equation}
The ratio of the two terms in equation~(\ref{fe}) then becomes
\begin{equation}
 {\Phi_{,tt}\over \nabla^2\Phi} \sim \dot{R}^2 =v_r^2.
\end{equation}
This suggests that $\Phi_{,tt}$ can be neglected, and hence that
the QE approximation can be applied, only when $v_r^2\ll1$.
In Figure~\ref{fig14} we show the maximum radial velocity for different
values of $R_i/M_0$, and also plot the ratio of 
$\Phi_{,tt}$ and $2 \Phi_{,r}/R$ (which scales with $\nabla^2 \Phi$,
but is finite at the shell).  This plot suggests that for $R_i
/M_0 > 1$ we should expect errors less than a few percent, while
for $R_i/M_0 = 1$, as in our ultra-strong field case, we should
expect errors on the order of 10 \%, which is consistent with our
results.


\section{Conclusion}
\label{conc}

In this paper, we illustrate our QE approach for calculating
gravitational waveforms for a model problem in scalar gravitation. We
demonstrate why it is a very viable technique to modeling binary
inspiral in GR.

Compact binaries, consisting of neutron stars or black holes, emit
gravitational wave and slowly spiral towards each other until they
reach the ISCO.  Outside of the ISCO, the inspiral is very slow: the
gravitational radiation reaction timescale is much longer than the
orbital period.  It is therefore reasonable to assume the binaries to
be in QE, and the inspiral to proceed along a sequence of QE
configurations.  In a recent paper, Duez, Baumgarte and
Shapiro~\cite{dbs} demonstrated how the inspiral can be modeled by
inserting QE binary models as matter sources in Einstein's field
equations.  Integrating Einstein's equations then yields the
gravitational wave luminosity, from which the inspiral rate and hence
the entire gravitational wavetrain can be constructed.

In this paper, we study a simple analogue in scalar gravitation.
Scalar gravitation has the advantage that it is conceptionally
simpler, and that it admits gravitational radiation even in spherical
symmetry, so that we can reproduce all qualitative characteristics of
the QE approximation even in a $1+1$ dimensional problem.  We study
the radiative damping of an oscillating, spherical matter shell.  In
analogy with the orbiting binary in GR, the oscillating shell emits
gravitational radiation, and hence looses energy.  While the binary
separation decreases in the process, the oscillation amplitude
decreases.  Ultimately, the oscillation is completely damped out, and
a static equilibrium configuration is reached.

The advantage of scalar gravitation is that it is possible to
integrate the exact equations without any approximation.  We compare
these exact results with predictions from both our QE approach and the
monopole approximation (which is the analogue of the quadrupole
approximation in GR).  We find that all three approaches agree very
well for weak field solutions, but that the QE approach reproduces the
exact solution much better for moderately strong fields.  Only for
ultra-strong fields does the QE approach break down.

We conclude that the QE approximation is a very promising approach
to computing the adiabatic inspiral of compact binaries.


\acknowledgments

We thank M. Duez for useful discussions.  This work was supported by
NSF Grant PHY 99-02833 and NASA Grant NAG 5-7152 at the University of
Illinois at Urbana-Champaign.  TWB gratefully acknowledges support
through a Fortner Fellowship.  HJY acknowledges the support of the
Academia Sinica, Taipei.

\appendix


\section{The correspondence between a turning point in the energy curve
	and static equilibrium}
\label{mine}

In this section we show that a turning point of the quasi-equilibrium
energy $E_{\rm QE}$ versus apocenter radius $R_{\rm ap}$, 
\begin{equation} \label{QE_A1}
E_{\rm QE} = M_0\tilde{u}^0 + {1\over2}R \Phi_{\rm sh}^2,
\end{equation}
along a sequence of constant rest mass and angular momentum, coincides
with an equilibrium configuration in which all particles are in
circular orbits of radius $R_C$.

We evaluate $E_{\rm QE}$ at apocenter radius $R_{\rm ap}$, where
$\tilde{u}_r = 0$.  Eq.~(\ref{u0}) then yields
\begin{equation}
(\tilde{u}^0)^2=e^{2\Phi_{\rm sh}}+{\tilde{u}^2_\phi\over R^2_{\rm ap}}.
\end{equation}
For a sequence of constant rest mass and angular momentum we therefore
find
\begin{equation}
{d (M_0\tilde{u}^0) \over dR_{\rm ap}}
=- R_{\rm ap} \Phi_{\rm sh} {d\Phi_{\rm sh}\over dR_{\rm ap}}
-{\tilde{u}^2_\phi\over\tilde{u}^0 M_0 R^3_{\rm ap}},\label{t1}
\end{equation}
where we have substituted eq.~(\ref{C1}) in the form
\begin{eqnarray}
\Phi_{\rm sh}=-{M_0 e^{2\Phi_{\rm sh}}\over\tilde{u}^0 R_{\rm ap}}
\end{eqnarray}
to get eqn.~(\ref{t1}).
The derivative of the second term in eq.~(\ref{QE_A1}) is
\begin{equation}\label{t2}
{d\over dR}({\Phi^2R_{\rm ap}\over2})=R_{\rm ap}\Phi_{\rm sh}
{d\Phi_{\rm sh}\over dR_{\rm ap}} + {\Phi_{\rm sh}^2\over2}.
\end{equation}
Combining eqs.~(\ref{t1}) and~(\ref{t2}) we now find
\begin{equation}\label{t3}
{dE_{\rm QE}\over dR_{\rm ap}}={\Phi_{\rm
sh}^2\over2}-{M_0\tilde{u}^2_\phi\over\tilde{u}^0R^3_{\rm ap}}.
\end{equation}
At a turning point $dE_{\rm QE}/dR_{\rm ap} = 0$, which implies
\begin{equation}\label{an2}
\tilde{u}^2_{\phi} = - \frac{e^{2\Phi_{\rm sh}}}{2} R_{\rm
S}^2\Phi_{\rm sh}.
\end{equation}
Equation~(\ref{an2}) is equivalent to the result for static spherical shells, 
eq.~(\ref{u_phi_1}).
A turning point in the quasi-equi\-li\-brium energy therefore implies
\begin{equation}
R_{\rm ap}=R_{\rm S},
\end{equation}
and hence coincides with a static equilibrium configuration.


\section{Constructing the continuous QE wavetrain}
\label{construct}

In this appendix we discuss the assembly of the QE solution and
waveform.  To do so, we construct QE solutions for a set of apocenter
radii.  We then evolve the gravitational fields dynamically in the
presence of a QE matter source (see eq.~(\ref{qsfe})) to determine the
gravitational wave form and luminosity for each apocenter radius.
Guided by the analytical Newtonian expression for the shape 
of the waveform (see
Sec.~\ref{newton_limit}), we parameterize the waveform 
to be able to match the computed periodic data at discrete radii
to a smooth function.
In this Appendix we describe this
parameterization together with a prescription for how the time
evolution of these parameters can be determined.  We illustrate our
approach in Sec.~\ref{ill}, where we focus on the phase of the
gravitational waves and its dependence on other parameters, and
we construct the waveform template in Sec.~\ref{proc}.

\subsection{Waveform phase versus time}
\label{ill}

Consider the simple example of an oscillator with constant period $P$ 
and constant wave amplitude $A$.  The waveform $\Lambda$ can then 
be found from the phase $\theta$, which in turn is given as a function
of time:
\begin{eqnarray}
 \theta &=&\displaystyle{2 \pi \over P}t, \label{b11} \\
 \Lambda&=&A\sin\theta.
\end{eqnarray}
In general, however, both $A$ and $P$ vary with time, in which case
the phase, now denoted by $\vt$ can be computed from an integration
\begin{equation}\label{qex}
\begin{array}{l}
\vt=\int\displaystyle{2\pi\over P(t)}dt,\\
\quad\\
\Lambda=A(t)\sin\vt.
\end{array}
\end{equation}

In some cases, it is convenient to introduce 
a phase angle $\vp$ such that
\begin{equation} \label{b12}
\vt = \theta + \vp,
\end{equation}
where $\theta$ is given by eq.~(\ref{b11}).  Taking a time derivative
of this equation, we find
\begin{equation}
{2\pi\over P}=\dot\vt = \dot\theta+\dot\vp={2\pi\over P}-
{2\pi t\over P^2}\dot P+\dot\vp,
\end{equation}
where the first equality follows from~(\ref{qex}), the second
from~(\ref{b12}), and the third from~(\ref{b11}).  Comparing the left
and right hand side, we find $\dot\vp=2\pi\dot Pt/P^2$.

The relation between the phase $\theta$ and time $t$ may be more
complicated than eq.~(\ref{b11}), as for example in the Newtonian
analytic solution for an oscillating shell, eq.~(\ref{elliptic}).  We
therefore allow $\theta=\theta(t,p_i)$ to be a function of time $t$
and a set of additional parameters $p_i(t)$, and generalize
eqs.~(\ref{qex}) to
\begin{equation}\label{samp}
\begin{array}{l}
\vt=\int\p_t\theta dt,\\
\quad\\
\Lambda=\Lambda(\vt,p_i).
\end{array}
\end{equation}
We again introduce the phase angle $\vp$
\begin{eqnarray}
\vt&=&\theta+\vp,\\
\p_t\theta=\dot\vt&=&\dot\theta+\dot\vp=\p_t\theta+\Sigma
\dot p_i\p_{p_i}\theta+\dot\vp,
\end{eqnarray}
and now find
\begin{equation}\label{paf}
\dot\vp=-\Sigma\dot p_i\p_{p_i}\theta.
\end{equation}
This expression determines how the phase angle evolves when $\theta$
depends on several parameters.  We will use this expression for the 
scalar wave model problem below.

\subsection{QE parameterization and construction of the wave template}
\label{proc}

As described in Sec.~\ref{QES}, we construct QE models for a set of
apocenter radii and insert these as matter sources into the field
equations as matter sources (eq.~(\ref{qsfe})).  Dynamically evolving
the gravitational fields will then yield the gravitational wave
luminosity and the gravitational wave form. Given a suitable
parameterization of the wave form, 
whose shape may be far from sinusoidal,
we can find a set of parameters for
each chosen value of the apocenter, and can then construct an
interpolating function which yields all parameters as smooth functions
of the apocenter (e.g.~Figs.~\ref{fig4}, \ref{fig7} and \ref{fig11}).
Combining the QE energy with the gravitational wave luminosity
(eq.~(\ref{chain})) yields the damping rate and the entire evolution
of the system. 
This enables us to express all parameters, and hence the waveform, 
as a continuous function of time.

A suitable wave from template can be constructed from the Newtonian
analytical solution (see Sec.~\ref{newton_limit}).  Inserting
eq.~(\ref{elliptic}) into~(\ref{waveamp}) yields
\begin{eqnarray}\label{wf}
 t & = & \displaystyle{P\over2\pi}(u+e\sin u), \\
\Lambda & = & {8\pi\over3}{M^2_0\over R_{\rm ap}}\left[{e\sin u\over 
aP(1+e\cos u)^3}\right]_{t-r}.
\end{eqnarray}
We found that the fitting of the computed waveform to the above function
in strong field configurations could
be improved by allowing the eccentricity $e$ in the two equations
in eq.~(\ref{elliptic}) to be different. This yields the template
\begin{eqnarray}
t&=&{P\over2\pi}(u+e_P\sin u),\label{qetime}\\
\Lambda&=&{A\sin u\over(1+e_A\cos u)^3}.\label{qewave}
\end{eqnarray}
We thus have four independent parameters $P$, $e_P$, $A$, $e_A$
which we choose to characterize the waveform.

We construct QE configurations for 
15 different apocenter radii $R_{\rm
ap}$ between the initial radius $R_i$ and the final circular radius
$R_f$.  At each value of $R_{\rm ap}$ we determine the four parameters
$P$, $e_P$, $A$, $e_A$ as well as the gravitational wave luminosity,
$\Delta E/P$.  We then construct interpolating functions, so that
these parameters are now given as smooth functions of $R_{\rm ap}$.
Given the wave luminosity $\Delta E/P$ and the QE energy $E_{\rm QE}$,
the damping rate $dR_{\rm ap}/dt$ can be computed from
eq.~(\ref{chain}).  The four parameters $P$, $e_P$, $A$, $e_A$,
the wave luminosity $\Delta E/P$ and the  damping rate $dR_{\rm ap}/dt$
are now all available as a function of apocenter radius $R_{\rm
ap}$.  We show our results for weak, moderately strong 
and ultra-strong field cases
in Figs.~\ref{fig4}, \ref{fig7} and \ref{fig11}.

As in Sec.~\ref{ill}, we allow for a phase shift by introducing
\begin{equation}
\vt=u+\vp,
\end{equation}
where $u$ satisfies eq.~(\ref{qetime}), and where, according to 
eq.~(\ref{paf}), the phase angle $\vp$ evolves according to
\begin{equation}
\dot\vp = -\dot e_P\p_{e_P}u-\dot P\p_P u={2\pi t\dot P+P^2\dot
e_P\sin u\over P^2(1+e_P\cos u)}.
\end{equation}
The wave form $\Lambda$ is now given in terms of $\vt$ by
\begin{equation} \label{finalwave}
\Lambda = {A\sin\vt\over(1+e_A\cos\vt)^3}.
\end{equation}
Here, the time derivative of the four parameters 
can be derived from the chain rule
\begin{equation}
\dot p_i\equiv{dp_i\over dR_{\rm ap}} {dR_{\rm   
ap}\over dt}.
\end{equation}
The continuous waveform $\Lambda$ can now be constructed 
by integrating the two equations
\begin{eqnarray}
\dot u&=&{2\pi-u\dot P-(e_P\dot P+P\dot e_P)\sin u\over P(1+e_P\cos u)},\\
\dot \vp&=&{2\pi t\dot P+P^2\dot e_P\sin u\over P^2(1+e_P\cos u)}.\\
\end{eqnarray}
The sum of the two yields $\vt$, which, together with $A$ and $e_A$,
can then be inserted into eq.~(\ref{finalwave}) to give the entire
gravitational wave train.


\section{Numerical algorithm for integrating the scalar wave equation}
\label{numet}

In this Appendix, we outline our finite difference scheme, which 
is based on that of Ref.~\cite{slst}, and explain how the jump 
condition~(\ref{nonsmooth}) is implemented in our code.

In spherical symmetry, the field Eq. (\ref{fe2}) is
\begin{equation}\label{nufe}
\Phi_{,tt}={1\over r^2}(r^2\Phi_{,r})_{,r}+4\pi T.
\end{equation}
We rewrite this second order equation as two equations which are
first order in time
\begin{equation}\label{numla}
 \begin{array}{l}
  \T[\Phi]=\la,\\
  \quad\\
  \T[\la]=\R[\Phi]+4\pi T,
 \end{array}
\end{equation}
where
\begin{eqnarray}
\T[Y]&\equiv&Y_{,t},\label{nutimefield}\\
\R[Y]&=&6[r^3Y_{,r^2}]_{,r^3}.\label{nuregular}
\end{eqnarray}
Here $Y$ denotes either $\Phi$ or $\la$. The Laplacian in the field
equation is written in the form of (\ref{nuregular}) to ensure
regularity of the finite-difference operator near the origin.
We implement a leapfrog scheme with a variable time step
to solve these equations, and finite difference the operators
$\T[Y]$ and $\R[Y]$ according to
\begin{eqnarray}
\T^n_{i+1/2}[Y]&=&{\Delta t_{n-1}\over\Delta t_n+\Delta t_{n-1}}
{Y^{n+1}_{i+1/2}-Y^n_{i+1/2}\over\Delta t_n}\nonumber\\
&&+{\Delta t_n\over\Delta t_n+\Delta t_{n-1}}
{Y^n_{i+1/2}-Y^{n-1}_{i+1/2}\over\Delta t_{n-1}},\label{TY}\\
\R^n_{i+1/2}[Y]&=&{6\over r^3_{i+1}-r^3_i}\left[r^3_{i+1}{Y^n_{i+3/2}
-Y^n_{i+1/2}\over r^2_{i+3/2}-r^2_{i+1/2}}\right.\nonumber\\
&&\qquad\qquad\qquad\left.-r^3_i{Y^n_{i+1/2}-Y^n_{i-1/2}\over 
r^2_{i+1/2}-r^2_{i-1/2}}\right], \label{RY}
\end{eqnarray}
where $\Delta t_n=t_{n+1}-t_n$. These operators are second-order
accurate in both space and time. At $r=r_{\rm max}$ we impose an
outgoing wave boundary condition
\begin{equation}\label{owbc}
(rY)_{,t}+(rY)_{,r}=0,
\end{equation}
where $Y$ is either $\Phi$ or $\la$. A second-order accurate finite
difference form of this equation is
\begin{eqnarray}
Y^{n+1}_{\imax+1/2}&=&{r_{\imax-1/2}\over
r_{\imax+1/2}}Y^n_{\imax-1/2}+
{1-\zeta\over1+\zeta}\left[Y^n_{\imax+1/2}\right.\nonumber\\
&&\qquad\qquad\quad\left.-{r_{\imax-1/2}\over
r_{\imax+1/2}}Y^{n+1}_{\imax-1/2}\right],
\end{eqnarray}
where
\begin{equation}
\zeta={\Delta t_n\over r_{\imax+1/2}-r_{\imax-1/2}}.
\end{equation}

The jump condition~(\ref{nonsmooth}) is easiest implemented by letting
the grid move with the matter shell, so that $R_{\rm sh} = r_{i_{\rm
sh}+1/2}$ at all times.  The jump condition~(\ref{nonsmooth}) can then
be discretized to
\begin{eqnarray}\label{cubicdiscon}
& & r^2_{i_{\rm sh}+1}{\Phi^n_{i_{\rm sh}+3/2} -\Phi^n_{i_{\rm
sh}+1/2}\over r_{i_{\rm sh}+3/2}-r_{i_{\rm sh}+1/2}}-r^2_{i_{\rm
sh}}{\Phi^n_{i_{\rm sh}+1/2}- \Phi^n_{i_{\rm sh}-1/2}\over r_{i_{\rm
sh}+1/2}-r_{i_{\rm sh}-1/2}} \nonumber \\
& & \qquad ={M_0\over\tilde{u}^0}e^{2\Phi^n_{i_{\rm sh}+1/2}}.
\end{eqnarray}
Eq.~(\ref{cubicdiscon}) is used (by using a rootfinder)
to derive the value of $\Phi^{n+1}_{i_{\rm sh}+1/2}$ with the known
values of $\Phi^{n+1}_{i_{\rm sh}+3/2}$ and $\Phi^{n+1}_{i_{\rm sh}-1/2}$
which are obtained from eqs.~(\ref{TY}) and (\ref{RY}) in advance.
After each time step, we
interpolate the function values (at all three timelevels $n+1$, $n$
and $n-1$) to a new grid, so that $r_{i_{\rm sh}+1/2}$ at the new time
level $n+1$ coincides with the new location of the matter shell
$R_{\rm sh}$.  In our runs, we have chosen $i_{\rm sh} = 50$.  We use
a uniform grid inside the shell, and a geometric progression,
i.e., $r_{i+3/2}-r_{i+1/2}=\alpha(r_{i+1/2}-r_{i-1/2})$
where $\alpha$ is a fixed ratio, near unity, in the exterior.

Since all the particles making up the shell are identical, it is
sufficient to evolve one particle.  Moreover, since
$\tilde{u}_\phi$ is a constant of the motion,
only eqs.~(\ref{r_t}) and (\ref{radiuseqn}) need to be integrated 
to determine the particle's geodesic motion.



\begin{thebibliography}{99}

\bibitem{kip}
 K. S. Thorne, in {\it Black Holes and Relativistic Stars}, 
edited by R.M. Wald (U. of Chicago Press, Chicago, 1998), p.~62.

\bibitem{djs00} T. Damour, P. Jaranowski and G. Sch\"afer, Phys. Rev. D
	{\bf 62}, 084011 (2000);
        L. Blanchet, T. Damour, B. R. Iyer, C. M. Will, and A. G. Wiseman,
        Phys. Rev. Lett {\bf74} 3515 (1995).

\bibitem{bcsst} T. W. Baumgarte, G. B. Cook, M. A. Scheel, S. L. Shapiro
	and S. A. Teukolsky, Phys. Rev. Lett. {\bf 79}, 1182 (1997);
	Phys. Rev. D {\bf 57}, 7299 (1998).

\bibitem{irrotational} S. Bonazzola, E. Gourgoulhon, and J. A. Marck, 
	Phys. Rev. Lett. {\bf 82}, 892 (1999); 
	P. Marronetti, G. J. Mathews and J. R. Wilson,
	Phys. Rev. D {\bf 60}, 087301 (1999);
 	K. Uryu and Y. Eriguchi,
	Phys. Rev. D {\bf 61}, 124023 (2000).
	K. Uryu, M. Shibata and  Y. Eriguchi, submitted (also gr-qc/0007042).

\bibitem{c94} G. B. Cook, Phys. Rev. D {\bf 50}, 5025 (1994);
	T. W. Baumgarte, Phys. Rev. D {\bf 62}, 024018 (2000).

\bibitem{stu}
        S. L. Shapiro, paper presented at the Binary Neutron Star 
        Grand Challenge Workshop, NCSA, Urbana, Il, January 1997.

\bibitem{bhs}
T. W. Baumgarte, S. A. Hughes and S. L. Shapiro, 
Phys. Rev. D {\bf60}, 087501 (1999).

\bibitem{dbs}
M. D. Duez, T. W. Baumgarte and S. L. Shapiro, submitted (also gr-qc/0009064).

\bibitem{mtw}
C. W. Misner, K. S. Thorne, and J. A. Wheeler, {\it Gravitation} 
(Freeman, San Francisco, 1973).

\bibitem{slst}
 S. L. Shapiro and S. A. Teukolsky, Phys. Rev. D {\bf47},1529 (1993) [ST].

\bibitem{slst2}
        S. L. Shapiro and S. A. Teukolsky, Phys. Rev. D {\bf49}, 
        1886 (1994); M. A. Scheel, S. L. Shapiro and S. A. Teukolsky,
        Phys. Rev. D, {\bf49}, 1894 (1994).

\bibitem{wm95} J. R. Wilson and G. J. Mathews, Phys. Rev. Lett. {\bf 75},
	4161 (1995).



\end{thebibliography}
\end{document}